# COSMOLOGICAL EVOLUTION AND LUMINOSITY FUNCTION EFFECTS ON NUMBER COUNTS, REDSHIFT AND TIME DILATION OF BURSTING SOURCES

A. Mészáros

Astronomy Dpt, Charles University, Švédská 8, 150 00 Prague 5, Czech Republic

and

P. Mészáros [1,2]

525 Davey Laboratory, Pennsylvania State University, University Park, PA 16803

## ABSTRACT

We present analytic formulae for the integral number count distribution of cosmological bursting or steady sources valid over the entire range of fluxes, including density evolution and either standard candle or a power law luminosity function. These are used to derive analytic formulae for the mean redshift, the time dilations and the dispersion of these quantities for sources within a given flux range for Friedmann models with $\Omega = 1$, $\Lambda = 0$ without K-corrections, and we discuss the extension to cases with $\Omega < 1$ and inclusion of K-corrections. Applications to the spatial distribution of cosmological gamma ray burst sources are discussed, both with and without an intrinsic energy stretching of the burst time profiles, and the implied ranges of redshift $z$ are considered for a specific time dilation signal value. The simultaneous consideration of time dilation information and of fits of the number distribution versus peak flux breaks the degeneracy inherent in the latter alone, allowing a unique determination of the density evolution index and the characteristic luminosity of the sources. For a reported time dilation signal of 2.25 and neglecting [including] energy stretching we find that the proper density should evolve more steeply with redshift than comoving constant, and the redshifts of the dimmest sources with stretching would be very large. However, the expected statistical dispersion in the redshifts is large, especially for power law luminosity functions, and remains compatible with that of distant quasars. For smaller time dilation values of 1.75 and 1.35 the redshifts are more compatible with conventional ideas about galaxy formation, and the evolution is closer to a comoving constant or a slower evolution. More generally, we have considered a wide range of possible measured time dilation ratios and discuss the values of the density evolution and the redshifts that would be expected for different values of the energy stretching.

---

[1] Institute for Theoretical Physics, University of California, Santa Barbara, CA 93106

[2] Center for Gravitational Physics and Geometry, Pennsylvania State University, University Park, PA 16803

*Subject headings:* cosmology — source counts — gamma-rays: bursts.

## 1. Introduction

The integral distribution $N(>F)$ of the number $N$ of sources with flux greater than $F$ provides valuable information about the luminosity and spatial distribution of unidentified astronomical sources, especially in the absence of independent distance indicators. The main problem is that it convolves the information about the luminosity, density and distance distributions in a manner which is difficult to untangle. For gamma ray burst (GRB) sources, the angular distribution is highly isotropic and appears consistent with either a cosmological or an extended galactic halo interpretation (Fishman, *et al.*, 1994). In either case, the departure of the integral distribution from a simple Euclidean $N(>F) \propto F^{-3/2}$ law observed at low fluxes may be an indication that the end of the spatial distribution has been reached (Meegan, *et al.*, 1992), and/or that at low fluxes the effects associated with the luminosity function or the density evolution start to dominate the integral distribution (e.g. Wasserman, 1992, Wijers & Paczyński, 1994). In the cosmological case, even an unbounded and unevolving standard candle distribution will slowly turn over at low $F$ due to cosmological redshift effects at $z \gtrsim$ unity, e.g. Mao & Paczyński, 1992, Dermer, 1992. However, such effects at low fluxes depend on the type of luminosity function and density evolution of the sources.

Information about the redshift of the sources would be of significant interest for a cosmological distribution, as it could fix one of the crucial quantities which is otherwise an unknown parameter in statistical $N(F)$ vs. $F$ fits (e.g. Loredo and Wasserman, 1992, Fenimore, *et al.*, 1993, Band, 1994, Horack, *et al.*, 1994, Emslie and Horack, 1994, Cohen and Piran, 1995, Mészáros & Mészáros, 1995, hereafter MM95). In the absence of identified counterparts and/or of identifiable lines, some information on the redshift could be obtained from the detection of a "cosmological signature", fainter (distant) sources being expected to have longer characteristic timescales due to cosmological time dilation (Paczyński, 1992, Piran, 1993). Evidence for this effect has been reported from BATSE 2B data (Norris, *et al.*, 1994, Norris, *et al.*, 1995, Fenimore & Bloom, 1995) but is not seen in some other analyses (e.g. Mitrofanov, *et al.*, 1994, Mitrofanov, *et al.*, 1995). One difficulty is that this effect is cleanest for standard candle sources with a standard duration; a broad luminosity function and/or an intrinsic spread in the durations could smear out the signature. Another possible difficulty with this signature is that its effects could be mimicked by intrinsic properties of the sources (e.g. Mészáros & Rees, 1993, Brainerd, 1994, Band, 1994, Yi, 1994). An additional complication is that an intrinsic energy stretching of the time profiles may be present (Fenimore & Bloom, 1995, Fenimore, *et al.*, 1995) which would also weaken the cosmological signature.

In this paper we present new exact analytic solutions for the integral distribution $N(>F)$

with a power law or standard candle luminosity function and a power density evolution law, in the context of cosmological distributions. These expressions are exact over the entire range of fluxes $F$, and thus are more general than the asymptotic forms discussed in MM95. We discuss mainly the $\Omega = 1$, $\Lambda = 0$ case without K-corrections (where $\Omega$ is the present ratio of the density to the critical one, $\Lambda$ is the cosmological constant), with comments on extensions to the cases $\Omega < 1$ and inclusion of K-corrections. We also discuss analytic expressions for the theoretically expected redshift and time dilation, calculating the mean values and the dispersion under the effect of both a density evolution and a luminosity function. The cosmological signature is sensitive to both, the main effect of a luminosity function being a reduction of the signature over what is expected in a standard candle model. Some analytic expressions are valid both for bursting or steady sources, with an appropriate change of the evolution index. The inclusion of an intrinsic energy stretching of time profiles in the example of GRB (such that bursts are intrinsically narrower at larger energies) is discussed and incorporated into an analysis of the redshifts and cosmological signatures of GRB. We model the expected signatures both with and without energy stretching effects for the evolution and luminosity function parameters, and assuming a given value for the cosmological signature illustrate how one can determine the density evolution index and deduce the range of redshifts for sources in different peak flux ranges. We also discuss the statistical errors associated with such redshift determinations, and discuss their compatibility with currently held views on the redshift of earliest galaxy formation.

## 2. Integral Distribution

In this section we derive analytic expressions for $N(>F)$ of bursting or steady sources. Note that the corresponding differential distribution used for statistical fits is obtainable as $N(F) = | (dN(>F)/dF) |$. The analytical relations collected in this section were already used (without equation details) in the $\chi^2$ fits to the 2B catalogue in Horváth, et al., 1995 (hereafter HMM95). Note that our calculations in §2 and §3 are general, and hold for any cosmological sources with $F$ in units of photons/(cm$^2$s). Note also that in this paper we denote the flux by $F$ (unlike in MM95 and HMM95). This more general notation is motivated by two things: first, our calculations in §2 and §3 may be applied in some cases to steady sources; and second, even for gamma ray bursts the expressions are valid either for average or peak photons fluxes, as long as $F$ is in units of photons/(cm$^2$s). We prefer $F$ rather than the $P$ sometimes used for peak flux, which can be confused with power or period, and this also distinguishes it from the instrumental counts/s, which are denoted generally by $C$. Thus, in §2 and §3 $F$ denotes generally photon number flux, in the remaining sections photon number peak flux of GRB, and $C$ are the instrumental counts/s. It is also necessary to emphasize that our value of $F$ is not to be confused with the *energy* flux measured e.g., in units ergs/(cm$^2$s). This is because the relation used here between the flux and luminosity (see below equation (3)) holds only in the case when the luminosity has the dimension photons/s, and the flux has dimension photons/(cm$^2$s). Otherwise, for energy fluxes our equation (3), and hence the following calculations, should be correspondingly changed (MM95).

We assume a density evolution as a power law of the scale factor,

$$n(z) = n_o(1+z)^D ,  \tag{1}$$

where $z$ is redshift, $D$ is a real number characterizing the density evolution, $n(z)$ is the source proper density, $n_o$ is the corresponding density at $z = 0$ ($D = 3$ corresponds to a constant comoving density), and the units of $n_o$ are in Mpc$^{-3}$yr$^{-1}$ for bursting sources or Mpc$^{-3}$ for steady sources. The luminosity function $\Phi$ is taken to be either of the standard candle type or a power law in the luminosity between some lower and upper limits in the photon luminosity $\mathcal{L}$ (in photon units s$^{-1}$),

$$\Phi(\mathcal{L}) = \begin{cases} n_o \delta(\mathcal{L} - \mathcal{L}_o) , & \text{(standard candle)} ; \\ \bar{n} \mathcal{L}_m^{-1} (\mathcal{L}/\mathcal{L}_m)^{-\beta} , & \text{for } \mathcal{L}_m \leq \mathcal{L} \leq \mathcal{L}_M \text{ (power law)} \end{cases} ,  \tag{2}$$

where $\bar{n} = n_o(\beta-1)/[1 - K^{-(\beta-1)}]$ for $\beta \neq 1$ and $K = \mathcal{L}_M/\mathcal{L}_m$. For $\Omega = 1$, if there is no K-correction (MM95), the flux $F$ is related to $\mathcal{L}$ and the comoving radial coordinate $\chi$ (Weinberg, 1972, MM95) by

$$\begin{aligned} F &= F_H (1-\chi)^2 \chi^{-2} , \quad F_H = \mathcal{L}/(4\pi R_o^2) \\ 1+z &= (1-\chi)^{-2} = \eta^{-2} , \end{aligned} \tag{3}$$

where $\chi = 1 - \eta$ varies between 0 and 1 for $z$ ranging from 0 to $\infty$ ($\eta$ is the conformal time), and where $R_o = (2c/H_o) = 6000 h^{-1}$Mpc is the Hubble radius ($c$ is the velocity of light; $H_o = 100h$ km/(s Mpc) is the Hubble parameter).

For a given density evolution index $D$ the number of observed sources with flux greater than $F$ (the integral distribution) is given by (MM95)

$$N_D(>F) = 4\pi R_o^3 \int_{\mathcal{L}_m}^{\mathcal{L}_M} \Phi(\mathcal{L}) d\mathcal{L} \int_0^{\chi_1} (1-\chi)^{6+2B-2D} \chi^2 d\chi , \tag{4}$$

where for bursting (steady) sources $B = 1$ ($B = 0$),

$$\chi_1 = (1 + [4\pi R_o^2 F/\mathcal{L}]^{1/2})^{-1} = (1 + \sigma^{1/2})^{-1} , \tag{5}$$

and we defined the normalized peak flux $\sigma = F/F_H$ as the ratio of the observed flux to the Euclidean flux from the same source at the Hubble radius. Concerning the parameter $B$, this is introduced in order to treat both steady ($B = 0$) and bursting sources ($B = 1$). The latter value arises because in this case one must include an extra $(1+z)^{-1}$ factor into equation (4) to account for time dilation of the burst rate per unit time, e.g. MM95, Mao & Paczyński, 1992. In what follows, we will consider bursting sources, i.e. $B = 1$. Nevertheless, the formulas presented in the entire §2 and §3.1 apply also for steady sources; for them one should use the appropriate dimension for steady sources and in §2 and §3.1 use the results based on equation (4) with $B = 1$, but substituting $D$ by $D + 1$. This means that what holds for a given $D$ in bursting sources also holds for $(D-1)$ in steady sources. In both cases, $D = 3$ means comoving constant source density.

Analytic solutions of equation (4) were given by MM95 for $F$ in the two (three) asymptotic regimes of the standard candle (power law) luminosity functions (2) and density evolution (1) in the special case of $D \leq 4$ with $D$ integer or half-integer. It is useful, however, to have analytic solutions which are valid for arbitrary $F$ and $D$. In the next two subsections we derive the full analytic solutions over the entire range of $F$ and $D$.

## 2.1. Standard Candle Luminosity Distribution

For the standard candle (SC) case in equation (4) we have $\int_0^\infty \Phi(\mathcal{L})d\mathcal{L} = n_o$, so the double-integral reduces to an integral over the single variable $\chi$. In MM95, the solution of the SC case was found by expanding the binomial in the integrand for integer or semi-integer values of the density evolution index $D \leq 4$. Here we do the direct integration of equation (4) in the SC case using $\eta$ as the variable of integration. The result is

$$
\begin{aligned}
N_D(> F) &= (4\pi/3) n_o R_o^3 \sigma_o^{-3/2} \, I_D(\sigma_o) \, , \\
\sigma_o^{-3/2} \, I_D(\sigma_o) &= 3 \Big( \frac{1 - \eta_1^{9-2D}}{9 - 2D} - \frac{1 - \eta_1^{10-2D}}{5 - D} + \frac{1 - \eta_1^{11-2D}}{11 - 2D} \Big) \, , \\
\lim_{F \to 0} N_D(> F) &= (4\pi/3) n_o R_o^3 A_D \, , \quad D < 4.5 \\
\lim_{F \to \infty} N_D(> F) &= (4\pi/3) n_o R_o^3 \sigma_o^{-3/2} \, , \\
A_D &= 6/[(9 - 2D)(10 - 2D)(11 - 2D)] \, ,
\end{aligned}
\qquad (6)
$$

where $I_D$ is a dimensionless function of $\sigma_o$ or $\eta_1$ which are defined as

$$
\sigma_o = F/F_{Ho} = F/[\mathcal{L}_o/(4\pi R_o^2)] \; ; \quad \eta_1 = 1 - \chi_1 = \Big[ 1 + \sigma_o^{-1/2} \Big]^{-1} \qquad (7)
$$

The expression of $N_D(> F)$ in (6) is of broader applicability than that in MM95, being valid for any real $D$ (except $D = 4.5, 5, 5.5$). (In MM95 only the two limiting cases were obtained for arbitrary $D \leq 4$, and the identical results in different forms for integer and semi-integer $D \leq 4$.)

For the three remaining values of $D$ the direct integration of equation (4) gives

$$
\sigma_o^{-3/2} I_{4.5} = -(9/2) + 6\eta_1 - 3 \ln \eta_1 - (3/2)\eta_1^2 \, , \qquad (8)
$$

$$
\sigma_o^{-3/2} I_5 = 3\eta_1^{-1} + 6 \ln \eta_1 - 3\eta_1 \, , \qquad (9)
$$

$$
\sigma_o^{-3/2} I_{5.5} = (9/2) - 3 \ln \eta_1 - 6\eta_1^{-1} + (3/2)\eta_1^{-2} \, . \qquad (10)
$$

Note that for $D \geq 4.5$ one has $\lim_{F \to 0} N_D(> F) = \infty$. Note also that for $D = 6$ we have

$$
N_6(> F) = (4\pi/3) n_o R_o^3 [\chi_1/(1 - \chi_1)]^3 = (4\pi/3) n_o R_o^3 \sigma_o^{-3/2} \, , \qquad (11)
$$

i.e., $I_6 = 1$. This is exactly the same expression as in the Euclidean limit for arbitrary $D$ (see equation (6)). For $D < 6$ ($D > 6$) the expressions $N_D(> F)$ grow less (more) steeply towards small $F$ than the corresponding Euclidean curve (11). In other words, $I_D < 1$ for $D < 6$, and $I_D > 1$ for $D > 6$. For example, for $D = 7$ we have

$$N_7(>F) = (4\pi/3)n_o R_o^3 \sigma_o^{-3/2}\left[1 + (3/2)\sigma_o^{-1/2} + (3/5)\sigma_o^{-1}\right] . \quad (12)$$

Note that the mimicking of an Euclidean behavior for an evolution law $D = 6$ is characteristic of an $\Omega = 1$, $\Lambda = 0$ bursting model. For instance, with the same model but steady sources, the Euclidean behavior occurs with $D = 5$, rather than 6. More generally, for any arbitrary cosmological model one can always find an evolution law that will just cancel out the cosmological effects.

In Figure 1 we show two sets of theoretical curves for two values of $\mathcal{L}_o$. For comparison, we also show the 2B observed integral numbers, with a sliding vertical axis. The top set of curves has an $\mathcal{L}_o$ chosen to give an approximate eye fit of the 2B data to the $D = 4$ curve, while the lower set has an $\mathcal{L}_o$ giving an approximate fit of the data to the $D = 2$ curve. While this is not an accurate fit, it illustrates the fact that fits to observed data can be found for various $D$ (including $D = 3$, not shown) by varying $\mathcal{L}_o$ (and $n_o$, but the vertical axis has been left arbitrary). More detailed $\chi^2$ joint fits to the 2B and PVO data sets give somewhat different $D, \mathcal{L}_o$ best fit values (see HMM95, and Table 1). Fits to 3B data are in progress. Qualitatively, this $D$ degeneracy can be understood by considering, e.g. what happens as one increases $\mathcal{L}_o$, in which case one needs to see deeper to maintain a given flux $F$. Seeing deeper gives more cosmological bending to $N(>F)$ for nonevolving bursts; to match the data one would then require an $N(>F)$ that bends less severely by incorporating a density evolution that increases the relative frequency of bursts at large distances.

### 2.2. Power Law Luminosity Distribution

The equation (4) for a power law (PL) luminosity function given by the second line of equation (2) reduces to the form (see MM95, equations (14)-(17))

$$N_D(>F) = \frac{4\pi \mathcal{L}_m^{3/2} \bar{n}}{3(4\pi F)^{3/2}} I_D = \frac{4\pi}{3} \bar{n} R_o^3 \sigma_m^{-3/2} I_D , \quad (13)$$

where

$$I_D = 2b^{-5+2\beta} \int_b^{bK^{1/2}} dy \, y^{4-2\beta}(1+y)^{-3}\left[1 + \sum_{k=1}^{8-2D} a_k y^k (1+y)^{-k}\right] , \quad (14)$$

and $I_D$ is again a dimensionless function of $\sigma_m$, $K$, $\beta$ and $D$, valid for integer and semi-integer values of $D \leq 4$ (for other values of $D$ see below). We have defined here

$$\begin{aligned}
y &= b(\mathcal{L}/\mathcal{L}_m)^{1/2}, \quad b = \sigma_m^{-1/2} = [\mathcal{L}_m/(4\pi R_o^2 F)]^{1/2}, \quad \sigma_m = F/F_{Hm} \\
F_{Hm} &= \mathcal{L}_m/(4\pi R_o^2), \quad F_{HM} = \mathcal{L}_M/(4\pi R_o^2), \quad K = \mathcal{L}_M/\mathcal{L}_m \geq 1, \\
a_k &= (-1)^k [3/(k+3)] \, [(8-2D)!/[k!(8-2D-k)!]], \quad k = 0, 1, ...., (8-2D) . 
\end{aligned} \quad (15)$$

$N_D$ is a function of $F$ ($\sigma_m$), and depends on the parameters $\bar{n}$, $\mathcal{L}_m$, $\mathcal{L}_M$, $\beta$ and $D$. In MM95 the integral (14) was estimated in three asymptotic regimes of $F$ for arbitrary $\beta$. However, exact solutions valid over the entire range of $F$ are of interest and in principle can be derived for any rational value of $\beta$. An exact solution is particularly simple for some definite rational values of $\beta$. Guided by the value $\beta \sim 1.88$, which fits the slope of the low $F$ differential distribution (e.g. Meegan, $et\ al.$, 1992), we adopt here $\beta = 15/8$. A nontrivial choice of $\beta$ must be near such a value, where the shape of the distribution at the faint end is dominated by the shape of the luminosity function, rather than cosmological effects. On the other hand, significantly steeper (e.g. $\beta \gtrsim 2$) power laws are dominated by the faint end, while significantly flatter power laws (e.g. $\beta \lesssim 1$) are dominated by the bright end, and thus mimic a standard candle behavior, the faint part of the distribution being dominated by cosmological effects. With this value of $\beta$, the expression for $I_D$ in the integer and semi-integer case $D \leq 4$ becomes

$$I_D = 8b^{-5/4} \int_{t_m}^{t_M} dt\ t^4 T^{-3} \sum_{k=0}^{M} a_k \left(t^4/T\right)^k = 8b^{-5/4} A_D \Psi_M \ , \qquad (16)$$

where

$$\begin{aligned} t &= y^{1/4} = b^{1/4}(\mathcal{L}/\mathcal{L}_m)^{1/8}\ , \quad T = 1 + t^4\ , \\ t_m &= b^{1/4} = \sigma_m^{-1/8}\ , \quad t_M = b^{1/4} K^{1/8} = (F_{HM}/F)^{1/8}\ ; \\ A_D &= 6/[(9-2D)(10-2D)(11-2D)]\ , \quad M = 8 - 2D\ , \end{aligned} \qquad (17)$$

and (see Appendix)

$$\Psi_M = A_D^{-1} \int_{t_m}^{t_M} dt\ t^4 T^{-3} \sum_{k=0}^{M} a_k(t^4/T)^k = \int_{t_m}^{t_M} dt\ t^4 T^{-3} \sum_{k=0}^{M} [(k+1)(k+2)/2] T^{-k}\ . \qquad (18)$$

The $\Psi_M$ have an exact solution over the entire range of $F$, and can be obtained recursively starting from $\Psi_0$ and $\Psi_1$ as shown in Appendix. The result for the indefinite integrals is

$$\begin{aligned} \psi_0(t) &= (3/8)(1/16\sqrt{2})\Theta(t) + (1/32)(t/T) - (1/8)(t/T^2)\ , \\ \psi_1(t) &= (1/4)(t^5/T^3) + (33/12)\psi_0(t)\ , \\ \psi_M(t) &= \left(\frac{M+1}{8}\right)\left(\frac{t^5}{T^{M+2}}\right) + \left(\frac{8M+3}{4M}\right)\psi_{M-1} - \left(\frac{4M+3}{4M}\right)\psi_{M-2}\ , \end{aligned} \qquad (19)$$

where the definite integrals are $\Psi_M = \psi_M(t_M) - \psi_M(t_m)$, and

$$\Theta(t) = \ln \frac{t^2 + \sqrt{2}t + 1}{t^2 - \sqrt{2}t + 1} + 2\arctan(\sqrt{2}t + 1) + 2\arctan(\sqrt{2}t - 1)\ . \qquad (20)$$

To get the asymptotic regimes, note that the function $A_D \psi_M(t) = \int dt\ t^4 T^{-3} \sum_{k=0}^{M} a_k(t^4/T)^k$ for $0 < t \ll 1$ may also be written as a power series of $t$. Its concrete form may be obtained by first using the formula for the sum of geometrical series, as well as $T^{-1} = \sum_{s=0}^{\infty}(-1)^s t^{4s}$, and

– 8 –

then doing the integration. This gives $A_D \psi_M(t) = (t^5/5) + ...$ (the integration constant is zero, because $\psi_M(0) = 0$), and for any $D$ the first term is the same. Then, if $t_M \ll 1$ is so small that for $t = t_M$ one may restrict oneself to this first term, one immediately obtains the Euclidean limit with $A_D \Psi_M \simeq t_M^5/5$. Hence, one reproduces the asymptotic $t \to 0$ ($F \to \infty$) behavior obtained in MM95 (see equation (18) in MM95). Similarly, for $t \gg 1$ the function $A_D \psi_M(t)$ may also be written as a series, if one uses $T^{-1} = t^{-4}(1 + t^{-4})^{-1} = t^{-4} \sum_{s=0}^{\infty} (-1)^s t^{-4s}$. Then, after integration, one has $A_D \psi_M(t) = A_D(const. - (7t^7)^{-1} + ...)$. (Here the integration constant is not zero, and is identical to $\lim_{t \to \infty} \psi_M(t)$, which is easily calculable using $\lim_{t \to \infty} \arctan(\sqrt{2}t \pm 1) = \pi/2$. Nevertheless, this integration constant is cancelled, once the definite integral $\Psi_M$ is calculated.) If $t_m \gg 1$ is so big that for $t = t_m$ one may restrict oneself to this first term, one reproduces the second asymptotic regime of MM95 for $t \to \infty$, $F \to 0$ (equation (24) of MM95), and one has $N_D(> 0) = (4\pi/3)n_o R_o^3 A_D$, as it must be. The third asymptotic behavior predicted in MM95 occurs for values of $K$ large enough that in some range of $F$ one may simultaneously approximate $t = t_m$ with the Euclidean limit, and $t = t_M$ with this second limit (see equation (20) of MM95). Thus, for large $K$ the formulae in this section also give the three asymptotic regimes described in MM95; but for $K \simeq 1$ the situation is similar to the SC case, as expected.

For the case $D > 4$, when $D$ is integer or semi-integer, in equation (13) $I_D$ are also directly calculable. First, one must solve the $\chi$ integrals in equation (4) directly, as in the SC case, and then one integrates over $\mathcal{L}$. The results are

$$\begin{aligned}
I_{4.5} &= b^{-5/4} \left[ (24/7)t^{-3} - (24/7)t^{-7} \ln T - 3tT^{-1} + (9/[28\sqrt{2}])\Theta(t) \right]_{t_m}^{t_M}, \\
I_5 &= b^{-5/4} \left[ -(48/7)t^{-3} + (48/7)t^{-7} \ln T + (6/[7\sqrt{2}])\Theta(t) \right]_{t_m}^{t_M}, \\
I_{5.5} &= b^{-5/4} \left[ 12t + (24/7)t^{-3} - (24/7)t^{-7} \ln T - (24/[7\sqrt{2}])\Theta(t) \right]_{t_m}^{t_M}, \\
I_6 &= (8/5)(K^{5/8} - 1), \\
I_{6.5} &= (8/5)(K^{5/8} - 1) + (3/4)(K^{9/8} - 1)\sigma_m^{-1/2}, \\
I_7 &= (8/5)(K^{5/8} - 1) + (4/3)(K^{9/8} - 1)\sigma_m^{-1/2} + (24/65)(K^{13/8} - 1)\sigma_m^{-1}.
\end{aligned} \qquad (21)$$

We see that $I_6$ does not depend on $\sigma_m$, and hence $N_6(> F) \propto \sigma_m^{3/2} \propto F^{-3/2}$. Thus for $D = 6$ we obtain the Euclidean case both for SC and PL luminosity function. In fact, one can easily show that for $D = 6$ one obtains the Euclidean case for arbitrary luminosity functions $\Phi(\mathcal{L})$. From equations (4) and (5) it follows for $D = 6$

$$N_6(> F) = (4\pi/3)(4\pi F)^{-3/2} \int_0^\infty \Phi(\mathcal{L})\mathcal{L}^{3/2} dL = const \times F^{-3/2}, \qquad (22)$$

where $const$ does not depend on $F$. Obviously, for any $\Phi(\mathcal{L})$ and $D > 6$ the expected integral distribution $N_D(> F)$ is even larger than the Euclidean value. As mentioned at the end of §2.1, the Euclidean behavior of $D = 6$ is for $\Omega = 1$, $\Lambda = 0$ and bursting sources. For arbitrary cosmological models, it is always possible to find an evolution law that cancels out the cosmological effects.

Note that in Section 2.2 we restricted ourselves to the integer and semi-integer values of $D$. Nevertheless, the analytical formulae of this Section may be generalized for any real $D$. The relations (4-5) hold for any real $D$. Then the key problem is to write down the definite integral

$$\int_0^{\chi_1} \chi^2(1-\chi)^{8-2D}d\chi = \int_0^{\chi_1} Q_D(\chi)d\chi. \qquad (23)$$

The primitive function of $Q_D(\chi)$ for non-integer $(8-2D)$ is obtainable in at least two different ways. First, one writes for $M = 8 - 2D$ the Taylor series

$$(1-\chi)^M = 1 + \sum_{k=1}^{\infty}(-1)^k[M(M-1)...(M-k+1)/k!]\chi^k, \qquad (24)$$

where $k$ is integer. This is not new for $M$ integer (see MM95), where the series is finite because for $k \geq M+1$ one obtains identically vanishing terms. On the other hand, for non-integer $M$ we have an infinite series. In any case, one should use the relation (5), and in this series one obtains integer exponents of $\mathcal{L}$. Then the integration over $\mathcal{L}$ is analytical (although it can be cumbersome) and in principle can be done for any rational $\beta$. Second, one substitutes $(1-\chi) = \eta$, and obtains a simple primitive function of $Q_D$ from integration over $\eta$; this primitive function is a sum of three components, containing either $\ln \eta_1$ or - in general non-integer - powers of $\eta_1$. Therefore here, for non-integer $M$, the technical problems can arise from the integration over $\mathcal{L}$. In other words, the simpler integration over $\chi$ may lead to a more complicated integration over $\mathcal{L}$. In this paper we do not discuss these technical questions any further, and restrict ourselves to the cases of integer $M = 8 - 2D$.

In Figure 2 we show two sets of theoretical PL curves, both for $K = \mathcal{L}_M/\mathcal{L}_m = 10^2$. The particular choices of $\mathcal{L}_M, \mathcal{L}_m$ used here are not unique, but have been selected so as to give an approximate eye fit of the 2B data set to the $D = 4$ curve (top group of curves), and to the $D = 2$ curve (lower group of curves). The visual match of these theoretical PL curves to the 2B data is as good as that for the SC case in Figure 1, even though here we have a luminosity ratio of 100. (In fact, detailed $\chi^2$ fits to joint 2B and PVO data sets with PL models of $K = 10^2$ are indistinguishable to within $1\sigma$ from corresponding SC fits, as shown in HMM95 (see also Hakkila, et al. , 1994); such detailed PL fits to both 2B and PVO lead to more accurate choices of $\mathcal{L}_M, D, K$ values, which have been used in Table 3). The approximate eye fits to 2B shown in Figure 2 merely illustrate the fact that in the PL case, different choices of $\mathcal{L}_M, \mathcal{L}_m$ (even for constant $K$) lead to a different $D$ which matches the same observed data set.

In Figure 3 we compare the generic shape of the SC curves for various $D$ (top set of curves) to that of the generic PL curves (lower set of curves). For the latter we have chosen an artificially large spread of luminosities $K = 10^5$ with maximum luminosity $\mathcal{L}_M$ equal to that of the SC luminosity $\mathcal{L}_o$. This large spread $K$ permits one to see more easily the three limiting asymptotic regimes of the PL curves, going from the Euclidean $F^{-3/2}$ to $F^{1-\beta}$ (where $\beta = 15/8$ is used) to $F \rightarrow$ constant for $D \leq 4$ as $F$ is decreased. The intermediate regime present in the PL case (e.g. Wasserman, 1992, MM95) is not present in the SC case, We note that the detailed $\chi^2$ fits of HMM95 indicate that there are statistically acceptable fits to the 2B data set alone with $K = 10^5$, but only to $K \lesssim 10^2$ if 2B plus PVO is used.

## 3. Redshifts and Time Dilation

### 3.1. Mean Redshifts and Dispersions

The redshift of a source of luminosity $\mathcal{L}$ producing a flux $F$ is, from (3), (5),

$$1 + z = [1 + (F/F_H)^{-1/2}]^2 \; ; \quad F_H = \mathcal{L}/[4\pi R_o^2] \; . \tag{25}$$

For a given $\Phi(\mathcal{L})$ one has

$$N_D(> F_1) = 4\pi R_o^3 \int_{\mathcal{L}_m}^{\mathcal{L}_M} \Phi(\mathcal{L}) d\mathcal{L} \int_0^{\chi_1} Q_D(\chi) d\chi \; , \tag{26}$$

where $Q_D(\chi) = (1 - \chi)^{8-2D} \chi^2$ and for the flux $F_1$ we have $\chi_1 = [1 + (F_1/F_H)^{1/2}]^{-1}$. The number of observed sources, that give fluxes between $F_1$ and $F_2$ is ($0 \leq F_1 < F_2 \leq \infty$, including $F_1 = 0$ and $F_2 = \infty$)

$$N_D(F_1, F_2) = 4\pi R_o^3 \int_{\mathcal{L}_m}^{\mathcal{L}_M} \Phi(\mathcal{L}) d\mathcal{L} \int_{\chi_1}^{\chi_2} Q_D(\chi) d\chi = N_D(> F_1) - N_D(> F_2) \; , \tag{27}$$

where $\chi_2 = [1 + (F_2/F_H)^{1/2}]^{-1}$. In this equation there is a double-integral of the function $\Phi(\mathcal{L}) Q_D(\chi)$ depending on variables $\chi$ and $\mathcal{L}$. If one takes the integral of the function $(1 - \chi^2)^{-2} \Phi(\mathcal{L}) Q_D(\chi)$ for the same area of the variables $\mathcal{L}$ and $\chi$, and divides the result by $N_D(F_1, F_2)$, obviously one obtains the average value of the quantity $(1 - \chi^2)^{-2}$. But since $(1 + z) = (1 - \chi)^{-2}$, this gives the average value of the scale factor $\overline{(1 + z)}$ for all the sources with fluxes between $F_1$ and $F_2$, namely

$$\overline{(1 + z(F_1, F_2))} = \frac{\int_{\mathcal{L}_m}^{\mathcal{L}_M} \Phi(\mathcal{L}) d\mathcal{L} \int_{\chi_1}^{\chi_2} (1 - \chi)^{-2} Q_D(\chi) d\chi}{\int_{\mathcal{L}_m}^{\mathcal{L}_M} \Phi(\mathcal{L}) d\mathcal{L} \int_{\chi_1}^{\chi_2} Q_D(\chi) d\chi} = \frac{N_{D+1}(F_1, F_2)}{N_D(F_1, F_2)} \; . \tag{28}$$

This expression follows from the relation $(1 - \chi)^{-2} Q_D(\chi) = Q_{D+1}(\chi)$ and is valid for arbitrary luminosity functions.

Also, since $Q_{D+2}(\chi) = (1 - \chi)^{-4} Q_D(\chi)$ and $(1 + z)^2 = (1 - \chi)^{-4}$, it follows similarly that

$$\overline{(1 + z(F_1, F_2))^2} = \frac{\int_{\mathcal{L}_m}^{\mathcal{L}_M} \Phi(\mathcal{L}) d\mathcal{L} \int_{\chi_1}^{\chi_2} (1 - \chi)^{-4} Q_D(\chi) d\chi}{\int_{\mathcal{L}_m}^{\mathcal{L}_M} \Phi(\mathcal{L}) d\mathcal{L} \int_{\chi_1}^{\chi_2} Q_D(\chi) d\chi} = \frac{N_{D+2}(F_1, F_2)}{N_D(F_1, F_2)} \; . \tag{29}$$

Then

$$\Delta(1 + z(F_1, F_2)) = \left[\overline{(1 + z(F_1, F_2))^2} - \overline{(1 + z(F_1, F_2))}^2\right]^{1/2} \tag{30}$$

characterizes the dispersion of $(1 + z)$ around the mean value. Again, it is worth mentioning that these specific expressions are valid for $\Omega = 1$, $\Lambda = 0$ models without K-corrections. However, similar expressions can be derived for arbitrary models.

Clearly, for a given $F_1$ and $F_2$ the values of $\overline{(1 + z(F_1, F_2))}$ and $\overline{(1 + z(F_1, F_2))^2}$ are different for different $D$. In the SC case they depend also on $\mathcal{L}_o$, and in the PL case on $\mathcal{L}_m$ and $K$, but they do not depend on $n_o$, because the latter parameter cancels out in the definitions.

As a consistency test of these relations consider a special case where $F_2 = (F_1 + dF)$, where $dF$ is infinitesimally small, and - in addition - we have an SC luminosity function. In this special case it follows that

$$\begin{aligned} N_D(F_1, (F_1 + dF)) &= |(dN_D(>F)/dF)|_{F=F_1}\, dF, \\ \overline{1 + z(F_1, (F_1 + dF))} &= (1 - \chi_1)^{-2}\ ;\quad \Delta(1 + z(F_1, (F_1 + dF))) = 0, \end{aligned} \quad (31)$$

where we used the relation $(dN_D(>F)/dF) = (dN_D(>F)/d\chi)(d\chi/dF)$. We see that in this special case, the mean redshift equals to the redshift defined by $\chi_1 = (1 + (F_1/F_H)^{1/2})^{-1}$, and there is no dispersion, as expected.

### 3.2. Time Dilations and Energy Stretching

While the previous sections were valid both for steady or bursting sources, from here on we specialize to the latter case. Another observable quantity of potential cosmological significance for bursting sources is the intrinsic duration of source events, e.g. the time interval between the beginning and end of a burst, or a subpulse of a burst, or a characteristic variation timescale of the source flux (or the inverse frequency of a spectral line, if such exists). This duration may be quite different for different energy bands.

In the simplest case the intrinsic duration is "grey", or independent of the energy band, e.g. if there is a "standard" GRB burst duration $\Delta t_o$ in the rest-frame of the source. Then the apparent duration of a source placed at redshift $z$ is $\Delta t = \Delta t_o(1 + z)$ (Paczyński, 1992, Piran, 1993), and two bursts placed at different redshifts $z_1$ and $z_2$ will show different apparent durations related by (Fenimore & Bloom, 1995)

$$\tau_{12} = (\Delta t_1/\Delta t_2) = (1 + z_1)/(1 + z_2)\ . \tag{32}$$

A less simple case is that where the intrinsic burst durations are not grey, i.e. there is an intrinsic energy stretching, such that the intrinsic duration of the event in its own rest frame depends on the photon energy or waveband. This is reported to be the case with GRB, the same burst exhibiting increasingly shorter durations at higher photon energies (Fenimore, et al. , 1995, Fenimore & Bloom, 1995). As argued by these authors, the BATSE 2B data indicates an energy dependence of the type $\Delta t_o(E) \simeq \Delta t_o(E_o)(E/E_o)^{-k}$, where $E_o$ is an arbitrary energy, $\Delta t_o(E_o)$ has the dimension of time, and $k \simeq 2/5$. Assuming this to be the case and - without loss of generality - taking $k$ to be an arbitrary real number, one can show that if bursts had some standard duration at the same energy in their own rest frame, then the result of observing from different redshifts would lead to an apparent duration-redshift relation given by

$$(\Delta t_1(E_o)/\Delta t_2(E_o)) = [(1+z_1)\Delta t_o(E_1)]/[(1+z_2)\Delta t_o(E_2)] = [(1+z_1)/(1+z_2)]^{1-k} , \qquad (33)$$

where $E_1 = (1+z_1)E_o$, $E_2 = E_o(1+z_2)$. Because the right-hand-side does not depend on $E_o$, the ratio of apparent durations is $[(1+z_1)/(1+z_2)]^{1-k}$ for arbitrary $E_o$, and this energy need not be specified later. Thus in general we can write

$$\tau_{12} = (\Delta t_1/\Delta t_2) = [(1+z_1)/(1+z_2)]^j = r_{12}^j , \qquad (34)$$

where $\tau_{12}$ is the time ratio, $r_{12}$ is the redshift factor ratio, $j = 1$ if there is no energy stretching, or $j = 1 - k$ if there is energy stretching.

If we take events at two different flux levels $F_d$ and $F_b$ ($F_b > F_d$), assumed to come from an SC luminosity function, these correspond to redshifts $z_d$, $z_b$ ($z_d > z_b$) which come from the relation

$$F_{b,d} = \mathcal{L}_o/[4\pi R_o^2(\sqrt{1+z_{b,d}} - 1)^2] , \qquad (35)$$

(valid for $\Omega = 1$, $\Lambda = 0$, no K-correction; see §5 for a relaxation of some of these assumptions). Then, if we measure the ratio of the corresponding event durations (assumed to be some standard duration in their respective rest frame), $\tau_{db} = (\Delta t_d/\Delta t_b) > 1$, we have

$$(1+z_d)/(1+z_b) = \tau_{db}^{1/j}, \quad F_b/F_d = [(\sqrt{1+z_d}-1)/(\sqrt{1+z_b}-1)]^2, \qquad (36)$$

and hence

$$\begin{aligned}
\sqrt{1+z_b} &= (\sqrt{F_b/F_d} - 1)/(\sqrt{F_b/F_d} - \tau_{db}^{1/(2j)}) , \\
\sqrt{1+z_d} &= \tau_{db}^{1/(2j)}\sqrt{1+z_b} .
\end{aligned} \qquad (37)$$

Thus in this special case, both relevant redshifts are obtainable directly analytically. We see that one must have $F_b/F_d > \tau_{db}^{1/j} > 1$, and as $F_b/F_d$ runs from $\tau_{db}^{1/j}$ to $\infty$, $z_b$ decreases from $\infty$ to zero, i.e larger $F_b/F_d$ leads to smaller $z_b$. On the other hand, for a fix $F_b/F_d$ a smaller $\tau_{db}^{1/j}$ gives smaller redshifts.

To improve the statistics, one would want to bin the bursting source flux into some appropriately defined ranges of "dim" and "bright" categories (taking only sources whose fluxes are inside the ranges $F_b \pm \Delta F_b$ and $F_d \pm \Delta F_d$ centering around some mean values of dim and bright fluxes $F_d$ and $F_b$). One can then use the formulas (28) derived above to connect the corresponding ratios of the mean durations (assumed to be standard, with or without intrinsic energy stretching) to the ratio of the mean redshifts,

$$\overline{(1+z_d)} = \overline{(1+z((F_d - \Delta F_d),(F_d + \Delta F_d)))} ,$$

$$\begin{aligned}
\overline{(1+z_b)} &= \overline{(1+z((F_b - \Delta F_b), (F_b + \Delta F_b)))}\,, \\
\Delta(1+z_d) &= \Delta(1+z((F_d - \Delta F_d), (F_d + \Delta F_d)))\,, \\
\Delta(1+z_b) &= \Delta(1+z((F_b - \Delta F_b), (F_b + \Delta F_b)))\,, \\
\overline{(1+z_d)}/\overline{(1+z_b)} &= r_{db} = (\overline{\Delta t_d}/\overline{\Delta t_b})^{1/j} = \tau_{db}^{1/j}\,.
\end{aligned} \qquad (38)$$

One can also calculate the dispersion in this last ratio of mean values by combining the relative dispersions (30) for both groups alone via

$$\frac{\Delta\left(\overline{(1+z_d)}/\overline{(1+z_b)}\right)}{\overline{(1+z_d)}/\overline{(1+z_b)}} = \frac{\Delta r_{db}}{r_{db}} = \sqrt{\left(\Delta(1+z_d)/\overline{(1+z_d)}\right)^2 + \left(\Delta(1+z_b)/\overline{(1+z_b)}\right)^2}\,. \qquad (39)$$

This can be done for different parameters and separately for SC and PL luminosity functions, the mean redshifts and dispersions being connected to the integral number expressions $N_D$ for each type of luminosity function via equations (28,30).

## 4. Application of Time Dilation Analysis

Attempts to detect a "cosmological signature" (Norris, et al., 1994, Norris, et al., 1995) in GRB have compared the mean duration $\overline{\Delta t}$ of events in the BATSE data base with peak fluxes corresponding to count rates within specific bands $C_{dd} \pm \Delta C_{dd}$, $C_d \pm \Delta C_d$ and $C_b \pm \Delta C_b$, where the subindices are "dd" for dimmest, "d" for dim and "b" for bright bursts. The values chosen by Norris, et al., 1994 are $(C_{dd} - \Delta C_{dd}) = 1.4$ Kct/s, $(C_{dd} + \Delta C_{dd}) = 2.4$ Kct/s, $(C_d - \Delta C_d) = 2.4$ Kct/s, $(C_d + \Delta C_d) = 4.5$ Kct/s, $(C_b - \Delta C_b) = 18$ Kct/s, $(C_b + \Delta C_b) = 250$ Kct/s, where Kct/s = kilocounts/ second. These count values can be roughly translated (based on averages of data from Nemiroff, 1994, private communication) into the following peak fluxes: 1.4 Kct/s $\to$ 0.45 photons/(cm$^2$s); 2.4 Kct/s $\to$ 0.70 photons/(cm$^2$s); 4.5 Kct/s $\to$ 1.15 photons/(cm$^2$s); 18 Kct/s $\to$ 5.0 photons/(cm$^2$s); 250 Kct/s $\to$ 45.0 photons/(cm$^2$s). In what follows we lump together the "dim" and "dimmest" category into a single "dim" one labeled by the index d' (standing for d+dd in the above categories). Hence we have

$$\begin{aligned}
F_{d'} - \Delta F_{d'} &\sim 0.45\,, \quad F_{d'} + \Delta F_{d'} \sim 1.15\,, \quad F_b - \Delta F_b \sim 5.0\,, \quad F_b + \Delta F_b \sim 45.00\,, \\
F_{d'} &\sim 0.80\,, \quad \Delta F_{d'} \sim 0.35\,, \quad F_b \sim 25.0\,, \quad \Delta F_b \sim 20.0
\end{aligned} \qquad (40)$$

in units of photons/(cm$^2$s), implying a ratio $F_b/F_{d'} \sim 31$. Note that these are approximate 2B bands, which may differ depending on the analysis and the data cuts, but they serve to illustrate the effects discussed. A more detailed discussion of these bands is presented in Horack, et al., 1995b, who use a ratio $F_b/F_{d'} \sim 21$.

In Norris, et al., 1994, Norris, et al., 1995 an observed ratio of durations $\tau_{d'b} \simeq 2.25$ was reported. More recent values are 1.75 (e.g. Norris, 1995b, Horack, et al., 1995b) and 1.35 (Fenimore, 1995b). A first quick and simple estimate of the implications of such a time dilation can be obtained using the analytic formulae based on the flux bin centroid values (37). This gives

$$\begin{aligned}
z_b &= 0.26 \,, & z_{d'} &= 1.83 \,, & \tau_{d'b} &= 2.25, \, j = 1 \,, \\
z_b &= 0.60 \,, & z_{d'} &= 5.20 \,, & \tau_{d'b} &= 2.25, \, j = 3/5 \,, \\
z_b &= 0.16 \,, & z_{d'} &= 1.01 \,, & \tau_{d'b} &= 1.75, \, j = 1 \,, \\
z_b &= 0.32 \,, & z_{d'} &= 2.34 \,, & \tau_{d'b} &= 1.75, \, j = 3/5 \,, \\
z_b &= 0.07 \,, & z_{d'} &= 0.44 \,, & \tau_{d'b} &= 1.35, \, j = 1 \,, \\
z_b &= 0.14 \,, & z_{d'} &= 0.87 \,, & \tau_{d'b} &= 1.35, \, j = 3/5.
\end{aligned} \quad (41)$$

We see that the presence of an energy-stretching has an essential impact on the values of relevant redshifts; taken at face value a time dilation signal of 2.25 could imply that the dim burst are at redshifts larger than the most distant known objects (see also Fenimore, *et al.* , 1995). On the other hand, the lower time dilation values of 1.75, and especially 1.35, give more reasonable dim+dimmest redshifts, even with energy stretching. However, such a simple estimate based on equations (37) neglects binning effects as well as luminosity function and evolution effects.

More reliable conclusions may be obtained by calculating $\overline{(1+z_{d'})}$, $\overline{(1+z_b)}$, their ratio, and the corresponding dispersions via the formulas (28,30,38,39), which specifically include luminosity function and density evolution effects via the $N_D$ entering the definitions of the these bin-averaged quantities. For a putative measured time dilation $\tau_{d'b}$ one can then search for the cases when the ratio $r_{d'b} = \overline{(1+z_{d'})}/\overline{(1+z_b)}$ is equal (for $j = 1$, no stretching) to $\tau_{d'b}$ (2.25, 1.75, 1.35 in the Norris, *et al.* , 1995, Norris, 1995b, Fenimore, 1995b cases), or else is equal (for $j = 3/5$, with stretching) to $\tau_{d'b}^{5/3}$ (or 3.86, 2.54, 1.65 in the same three cases just mentioned). To do this we need to use the appropriate pairs of corresponding values $\mathcal{L}_o$ and $D$ (for SC) or triplets of corresponding values $L_m$, $K$ and $D$ (for PL) which give good $\log N - \log F$ fits (HMM95). We can restrict ourselves to values of $D \leq 5$, since $D \geq 6$ is excluded by the downward curvature of the counts at low $F$, and $D = 5.5$ gives also wrong fits (HMM95). The values calculated from our analytic expressions (38,39,28,30) using the SC values of $N_D$ from §2.1 are shown in Table 1.

These redshift estimates are systematically larger than the values from (41). Nevertheless, the dispersions of the average redshifts and their ratios are large, and therefore as a rough first approximation the values from (41) are compatible with those of Table 1. For instance, even in the case of $\tau_{d'b} = 2.25$, subtracting a $1\sigma$ dispersion from the mean value of $r_{d'b}$ for $D = 5$ gives an $r_{d'b}$ comparable to the mean $r_{d'b}$ value for $D = 4.5$, corresponding to $\overline{(1+z_{d'})} - 1 \sim 5.5$. This is near the redshift of the most distant quasars known.

The uncertainties in the mean redshifts are increased also by the fact that the estimates depend also on $\mathcal{L}_o$. The values used for each $D$ are from $\chi^2$ fits to the BATSE 2B data by HMM95, and a $\pm 1\sigma$ variation around the best fit can change $\mathcal{L}_o$ by factors $\sim 2$. In Table 2 we illustrate the impact of such $\lesssim 1\sigma$ changes in $\mathcal{L}_o$ for a fixed $D = 4.5$. We see that the impact on the redshifts of varying between different allowed $\mathcal{L}_o$ for the same $D$ is non-negligible. Nevertheless, while keeping such uncertainties in mind, it seems clear that if the time dilation of 2.25 (Norris, et al., 1995) were correct, under the SC assumption Table 1 indicates $\overline{z_b} \simeq (0.2 - 0.6)$ and $\overline{z_{d'}} \simeq (1.5 - 2.5)$ for $j = 1$ (no energy stretching), and $\overline{z_b} \simeq (0.6 - 2.0)$ and $\overline{z_{d'}} \simeq (4 - 10)$ for $j = 3/5$ (with energy stretching). For this time dilation, extremely large redshifts seem to be indicated for the dim objects (see also Fenimore, et al., 1995, Horack, et al., 1995b), but because of the dispersion around the mean values, even for $\tau_{d'b} = 2.25$ one cannot conclude necessarily that $z_{d'} > 5$ is required. For the lower values $\tau_{d'b} = 1.75, 1.35$ the dim redshifts are of course more moderate. This is discussed further in §6.

For the power law (PL) distribution luminosity function, we can use the same formalism to calculate the mean redshifts and their dispersions. We choose a relatively broad luminosity dispersion $K = (\mathcal{L}_M/\mathcal{L}_m) = 100$, which while still giving a good $\chi^2$ fit to the 2B+PVO data (HMM95), accentuates the differences between the SC and PL luminosity functions. While formally one uses the same equations (38, 39, 28, 30), the values of $N_D$ entering into them are now obtained from an integration over luminosities placed at different redshifts (§2.2). The results are summarized in Table 3.

These values appear slightly larger than in the SC case, but no far reaching conclusions may be drawn from this difference. In fact, the remarkable difference between Table 3 and Table 1 is provided by the much larger redshift dispersions for the power law luminosity distribution, as opposed to the standard candle case. Because of this large dispersion, the PL and SC results are clearly compatible with each other. Also, this very large dispersion implies that even for $\tau_{d'b} = 2.25$ the dimmest redshifts encompass within their $1\sigma$ error bars the values $z_{d'} \sim 1$, and the same true to a larger extent for the lower values of $\tau_{d'b}$. The larger dispersion of mean redshifts in Table 3 is quite reasonable, because for SC at a given $F$ one has a given $z$, but for PL at a given $F$ there is range of $z$ corresponding to a range of $\mathcal{L}$. The implications of such large dispersions for cosmological models for a given measured time dilation are further discussed in §6.

The dependence on $\mathcal{L}_m$ in the PL case is illustrated in Table 4. We see that, similarly to the SC case, the dependence on $\mathcal{L}_m$ is non-negligible, and while not changing the means by much it can change the ratios and further increase the dispersion. The dependence on $K$ is illustrated in Table 5. Here $D$ and $\mathcal{L}_m$ are fixed, and $K$ varies. In this case the $r_{d'b}$ practically remains unchanged, but the mean redshifts and dispersions vary significantly.

## 5. K-Correction and $\Omega < 1$ Effects

In the previous sections the calculations of the redshifts of GRBs or other sources was done without any K-corrections and for $\Omega = 1$. In this section the impact of relaxing these assumptions is estimated (c.f. also Fenimore & Bloom, 1995, Horack, et al., 1995b). For simplicity, we illustrate these effects for the special case when equations (36, 37) hold, since it was found that qualitatively they reproduce the behavior found in the more detailed analysis. We further simplify the situation by assuming that only one or the other effect operates alone, i.e., we either have $\Omega = 1$ and a K-correction, or $\Omega < 1$ and no K-correction.

If $\mathcal{L}_\nu \sim \nu^{\alpha-2}$, where $\mathcal{L}_\nu d\nu$ is the peak-flux luminosity in the frequency interval $\nu, (\nu + d\nu)$ and $\alpha$ is a real number, then there is no K-correction for $\alpha = 1$ (e.g. MM95). If $\alpha \neq 1$ then the peak-flux $F$ from a source at redshift $z$ will be $(1 + z)^{\alpha-1}$ times greater than the expected peak flux when no K-correction occurs. (For instance, for $\alpha = 0$ and $\mathcal{L}_\nu \sim \nu^{-2}$ one will have a $(1+z)$ times smaller peak flux than for $\alpha = 1$.) In other words, instead of equation (3) here one has:

$$F = \mathcal{L}/[4\pi R_o^2 (1+z)^{1-\alpha}(\sqrt{1+z}-1)^2]. \tag{42}$$

Assuming two sources with the same intrinsic peak flux spectra characterized by $\alpha$ at redshifts $z_b$ and $z_{d'}$, instead of equations (36, 37) one obtains

$$(1+z_{d'})/(1+z_b) = \tau_{d'b}^{1/j}, \quad F_b(1+z_b)^{1-\alpha}/[F_{d'}(1+z_{d'})^{1-\alpha}] = [(\sqrt{1+z_{d'}}-1)/(\sqrt{1+z_b}-1)]^2, \tag{43}$$

and hence

$$\sqrt{1+z_b} = (\sqrt{F_b \tau_{d'b}^{(\alpha-1)/j}/F_{d'}}-1)/(\sqrt{F_b \tau_{d'b}^{(\alpha-1)/j}/F_{d'}}-\tau_{d'b}^{1/(2j)}), \quad \sqrt{1+z_{d'}} = \tau_{d'b}^{1/(2j)}\sqrt{1+z_b}. \tag{44}$$

The impact of $\alpha \neq 1$ can be seen from the fact that these equations are identical to equations (36, 37), if one formally substitutes $F_b/F_{d'}$ by $F_b \tau_{d'b}^{(\alpha-1)/j}/F_{d'}$. Hence we see that for $\alpha < 1$ ($\alpha > 1$) one obtains systematically smaller (bigger) ratios for the peak fluxes, and hence bigger (smaller) redshifts than in the absence of a K-correction. To illustrate this effect consider again $F_b = 25$ and $F_{d'} = 0.8$ (in units of photons/(cm$^2$s) ). For $\alpha = 1$ we had (equation (41)) $z_b = 0.6$ and $z_{d'} = 5.2$ for $j = 0.6$. For the same case, but with $\alpha = 0$ (i.e. with $\mathcal{L}_\nu \sim \nu^{-2}$), we obtain $z_b = 1.1$ and $z_{d'} = 7.2$. This means that one may expect that K-correction effects, when relevant, will increase the redshifts that one derives. Generally $\alpha = 1$ is a good fit for the peak flux spectra of GRB in the range where most photons are collected from (e.g. Band, 1994), and therefore K-corrections may not play a large role for most sources. Nevertheless, one cannot exclude the situation where a systematically smaller $\alpha$ may be needed (e.g. Schaefer, et al., 1994) where K-corrections are relevant.

For $z \gtrsim$ few a value of $\Omega \neq 1$ may also affect the estimates of the relevant redshifts. We restrict ourselves to $\Omega < 1$. For $\alpha = 1$ we will have for a source with peak luminosity $\mathcal{L}$ at redshift $z$ a peak flux

$$F = \mathcal{L}/[4\pi(c/(H\sqrt{1-\Omega}))^2 \sinh^2 \chi (1+z)]. \tag{45}$$

In the denominator we again have $4\pi d_p^2(1+z)$, where $d_p$ is the proper distance (Weinberg 1972, MM95). The relation between $\chi$ and $1+z$ is now given by

$$1+z = (\cosh \eta_o - 1)/(\cosh(\eta_o - \chi) - 1), \tag{46}$$

– 17 –

where $\cosh \eta_o = (2/\Omega) - 1$. Hence

$$\sinh \chi = \frac{2\sqrt{1-\Omega}}{\Omega^2 (1+z)} \Big( 2(1-\Omega) + \Omega(1+z) \pm (2-\Omega)\sqrt{1-\Omega+\Omega(1+z)} \Big). \qquad (47)$$

To decide the right sign we note that for $z \to 0$ one should have $(c/(H\sqrt{1-\Omega}))\sinh \chi = cz/H$. This is fulfilled for the minus sign. Hence here, instead of equation (37), one obtains

$$\sqrt{\frac{F_b(1+z_{d'})}{F_{d'}(1+z_b)}} = \frac{2(1-\Omega) + \Omega(1+z_{d'}) - (2-\Omega)\sqrt{1-\Omega+\Omega(1+z_{d'})}}{2(1-\Omega) + \Omega(1+z_b) - (2-\Omega)\sqrt{1-\Omega+\Omega(1+z_b)}},$$

$$(1+z_{d'})/(1+z_b) = \tau_{d'b}^{1/j}. \qquad (48)$$

The system is solvable analytically for a given $\tau_{d'b}^{1/j}, \Omega$ and $F_b/F_{d'}$ (e.g., substituting $(1+z_{d'})$ from the second equation into the first one, a fourth order algebraic equation is obtained for the unknown $(1+z_b)$.) Nevertheless, it is much simpler to solve it numerically. In order to illustrate the trend, we solve it for $\tau_{d'b} = 2.25$, $j = 0.6$, $(F_b/F_{d'}) = 25$, for various $\Omega$. The relevant $z_b$ and $z_{d'}$ are summarized in Table 6. As seen from Table 6, decreasing values of $\Omega$ increase the relevant redshifts, as expected. More specifically, the numerical examples calculated here show that for the smallest values of $\Omega \simeq 0.2$ currently thought to be acceptable, an increase in redshift by up to a factor 2 may occur over the corresponding $\Omega = 1$ values.

## 6. Discussion and Conclusions

We have derived analytical expressions for the integral distribution of bursting or steady sources, as well as the mean redshifts, time dilations and the dispersions of these quantities over finite flux bandwidths, valid over the entire range of fluxes in a spatially flat Friedmann model. We have also evaluated the effects associated with $\Omega < 1$ and with the inclusion of K-corrections. These expressions are particularly useful for comparing against the numerical results $\log N - \log F$ fits, e.g. GRB catalogues such as the BATSE and PVO samples, and allow one to derive relatively simple and quick results without lengthy computations.

We have used our analytic expressions for the average redshifts in specific peak flux bands together with $\log N - \log F$ fits to the BATSE and PVO data on GRB to derive new information, based on measurements of time dilation effects. While the detection of such cosmological time dilation signals is currently the subject of debate (Norris, et al. , 1995, Fenimore, et al. , 1995, Mitrofanov, et al. , 1994, Mitrofanov, et al. , 1995), there is no doubt that under the cosmological interpretation such effects should be incorporated into analyses of the integral distribution in order to obtain more reliable conclusions about the redshift, luminosity and density behavior. If a measured time dilation exists, several important problems can be addressed. First, the ambiguities of previous $\log N - \log F$ fits (which allowed various choices of luminosities and of the density evolution index $D$, e.g. HMM95) are resolved. The inclusion of redshift information via a time dilation signal breaks the degeneracy of the models fits and allows a unique determination of $\mathcal{L}$ and $D$. Second, the use of our analytical redshift expressions using parameters from $\log N - \log F$ fits provide an estimate of the mean source redshifts in different peak flux bands that are explicitly consistent with a particular $\log N - \log F$ fit. Third, the same analytic expressions provide an explicit estimate of the dispersion (or essentially $1\sigma$ error bars) associated with these mean redshifts. This dispersion is especially large for the examples of a power law luminosity function, which provide good $\log N - \log F$ fits in a $\chi^2$ sense that are statistically indistinguishable from those of standard candles.

We have also shown that a simple analytic approximation to the Fenimore, *et al.*, 1995 energy stretching phenomenon of GRB time pulses implies that the *measured* time dilation factors between two arbitrary GRB flux classes labeled by $d''$ and $b$ would scale with the redshift factor approximately as $\tau_{d''b} = (\Delta t_d)/(\Delta t_b) = r_{d''b}^{3/5} = [(1 + z_{d'})/(1 + z_b)]^{3/5}$. This is a particularly simple scaling that allows us to find theoretical estimates of the bright and dim mean redshifts of bright and dim GRBs, if a time dilation signal $\tau_{d''b}$ is identifiable as being due to purely cosmological effect. Thus, for the value $\tau_{d''b} = 2.25$ cited by Norris, *et al.*, 1995, we find that the redshift factor ratio would the have to be $r_{d''b} = (1 + z_{d'})/(1 + z_b) = \tau_{d''b}^{5/3} \simeq 3.86$, if energy stretching is included. The numerical examples above are for a stretching $\Delta t \propto E^{-k}$ with a specific value (Fenimore, *et al.*, 1995) $k = 0.4$; we have in the text also provided general expressions for an arbitrary stretching index $k$.

As specific examples, we have discussed the redshifts implied by reported values of the time dilation $\tau_{d''b} = (\Delta t_{d'})/(\Delta t_b) \simeq 2.25$, 1.75, 1.35 (Norris, *et al.*, 1995, Norris, 1995b, Horack, *et al.*, 1995b, Fenimore, 1995b), under the assumption that these real and entirely cosmological, i.e. not contaminated by intrinsic effects (e.g. internal source physics which might mimic such a signature, etc.). Approximate standard candle (SC) results were obtained used the simple analytic equations 37, and more accurate results were obtained using the detailed analytic averaging over flux bands corresponding to bright and dim+dimmest bursts. The latter are detailed in Table 1 and Figure 4, from which one can find the relation between a particular redshift factor ratio $r_{d''b}$ and the density evolution factor $D$ (corresponding to a luminosity which gives consistency with the $\log N - \log F$ constraints). Table 1 and Figure 4 show that for $\tau_{d''b} = 2.25$, $r_{d''b} \simeq 2.25$ (no stretching) occurs for $D \simeq 4$, where $\overline{z_b} \simeq 0.5$ and $\overline{z_{d'}} \simeq 2.3$, while $r_{d''b} = 3.86$ (stretching) occurs for $D \simeq 5$, where $\overline{z_b} \simeq 1.6$ and $\overline{z_{d'}} \simeq 10.0$. For $\tau_{d''b} = 1.75$ and no stretching one infers $D \simeq 3$ and $\overline{z_b} \simeq 0.25$, $\overline{z_{d'}} \simeq 1.25$, while with stretching ($r_{d''b} = 2.54$) one gets $D \sim 4$ and $\overline{z_b} \simeq 0.5$, $\overline{z_{d'}} \simeq 1.5$. For $\tau_{d''b} = 1.35$ and no stretching the implied value is $D \sim 1$, $\overline{z_b} \simeq 0.1$ and $\overline{z_{d'}} \simeq 0.7$, while with stretching ($r_{d''b} = 1.65$) it is $D \simeq 2.5$, $\overline{z_b} \simeq 0.25$ and $\overline{z_{d'}} \simeq 1.15$.

The same three values of a putative time dilation $\tau_{d''b} = (\Delta t_{d'})/(\Delta t_b) \simeq 2.25$, 1.75, 1.35 were also used for comparing with the band-averaged redshift factor ratios in the power-law (PL) luminosity function case, detailed in Table 3 and Figure 5. From these one sees that the average redshifts in the PL luminosity function case are not significantly different from those obtained in the SC case case. For $\tau_{d'b} = 2.25$ without stretching a value of $r_{d'b} \simeq 2.25$ would occur for $D \simeq 4$, where $\overline{z_b} \simeq 0.9$ and $\overline{z_{d'}} \simeq 3.0$; while with with stretching a value of $r_{d'b} = 3.86$ occurs for $D \simeq 5$, where $\overline{z_b} \simeq 2.0$ and $\overline{z_{d'}} \simeq 10.0$. For $\tau_{d'b} = 1.75$ (no stretching) we have $D \simeq 3.5$, $\overline{z_b} \simeq 0.6$ and $\overline{z_{d'}} \simeq 1.8$; with stretching ($r_{d'b} = 2.54$) we have $D \simeq 4.5$ and $\overline{z_b} \simeq 1.4$ and $\overline{z_{d'}} \simeq 5.3$. For $\tau_{d'b} = 1.35$ (no stretching) we have $D \simeq 2$, $\overline{z_b} \simeq 0.3$ and $\overline{z_{d'}} \simeq 0.85$; with stretching ($r_{d'b} = 1.65$) we have $D \simeq 3.4$ and $\overline{z_b} \simeq 0.6$ and $\overline{z_{d'}} \simeq 1.7$.

For the standard candle case without the energy stretching of Fenimore, *et al.*, 1995, our theoretical model fits to the data give mean redshifts of bright and dim sources. We find from our SC/no stretch fits that for $\tau_{d'b} = 2.25, 1.75, 1.35$ the density evolution index required (see equation (1)) must be $D \sim 4, 3$ and $1$. With energy stretching, they would be $D \sim 5, 4$ and $2.5$. The large $D$ for $\tau_{d'b} = 2.25$ is in qualitative agreement with the results of Fenimore & Bloom, 1995, Fenimore, *et al.*, 1995, Horack, *et al.*, 1995b in independent and complementary analyses. In the power law (PL) luminosity function, where we used a ratio of brightest to faintest intrinsic luminosity $K = 10^2$, for the same three values of $\tau_{d'b}$ the no stretch fits are $D \sim 4, 3.5$ and $2$, while with stretching they are $D \sim 5, 4.5$ and $3.4$. The source luminosities corresponding to these density evolution exponents $D$ (compatible with the $\log N - \log F$ constraints) are given in HMM95. Here $D > 3$ means more sources at large redshifts and $D < 3$ means more at small redshifts, since in our proper density notation $D = 3$ corresponds to comoving constant density. A confirmation of a clear cosmological time dilation would be needed to obtain stronger constraints on the density evolution.

We note that the redshift values and evolution indices quoted are based on a statistical average over specific flux ranges, giving mean redshift values which are systematically larger than the values (41) based on the simple analytic estimates of equations (37) for the centroid of the flux range. Nevertheless, the dispersions of the average redshifts and their ratios are large, and therefore as a rough first approximation the values from (41) are compatible with those of Table 1 for the SC case. For instance, subtracting a $1\sigma$ dispersion from the mean value of $r_{d'b}$ for $D = 5$ gives an $r_{d'b}$ comparable to the mean $r_{d'b}$ value for $D = 4.5$, corresponding to mean dim redshifts $\overline{z_{d'}} \sim 5.5$, if $\tau_{d'b} = 2.25$. This is near the redshift of the most distant quasars known, and thus even for this most extreme time dilation in the SC case there need not be incompatibility between the Norris, *et al.*, 1995 signal and the earliest galaxy formation redshifts.

The tendencies outlined above are stronger in the case of a power law (PL) luminosity distribution. While the mean values obtained are comparable, the redshift dispersion is understandably larger than in the SC case. For a relatively large ratio of maximum to minimum luminosities $K = 10^2$, which still give good $\log N - \log F$ fits, the PL dim and bright mean redshifts without or with energy stretching are somewhat larger than those in the corresponding SC cases given above. In the PL case, however, while again the mean dim redshifts are very large for a dilation signal of 2.25, whether stretching is included or not, the dispersion is so large that a dim redshift as low as $z_{d'} \sim 1$ is within the $1\sigma$ error bars of the mean values.

The smaller time dilation signals $\tau_{d'b}$ (e.g. 1.75 or 1.35) of course bring the mean redshift values closer to $z_{d'} \lesssim 1$. There is also the additional possibility that the purely cosmological part of a time dilation signal be even smaller than what is directly measured, in that some specific models already imply a fairly generic source-intrinsic time dilation signal. As an example, the dissipative fireball external shock spectra calculated by Mészáros, Rees and Papathanassiou, 1994 show that shorter duration bursts (higher $\Gamma$) also have higher intrinsic fluences (because from general physical arguments one finds that the radiative efficiency generally increases with $\Gamma$), e.g. as illustrated in their Figure 4 of that paper. This mimics a cosmological time dilation, it is not strongly model dependent, and is independent of redshift (see also Brainerd, 1994, Yi, 1994).

The specific numbers discussed above are for an $\Omega = 1$, $\Lambda = 0$, K-correctionless model. For an $\Omega < 1$ universe, or in cases where a K-correction is necessary, the mean redshifts are generally larger than in the $\Omega = 1$ uncorrected case, but the error bars remain large enough that even time dilations $\sim 2$ can remain compatible in the PL case with modest dim source redshifts.

In conclusion, we have derived analytic expressions for the integral number of sources per unit flux in an $\Omega = 1$, $\Lambda = 0$ Friedmann model with density evolution and either standard candle or power law luminosity functions, of general applicability to steady or bursting sources. We have shown how the density evolution affects the luminosities inferred for the sources, and have discussed the effects of a power law luminosity function on $\log N - \log F$ distributions. Using these expressions we have also derived the mean redshift and time dilation factor distributions over given finite flux bands, and the dispersions associated with them. We have shown that a time dilation signal of 2.25 (Norris, *et al.*, 1995), if purely cosmological, would imply values for the redshift of the dimmest bursts which are very large especially if one includes pulse energy stretching (Fenimore, *et al.*, 1995). However the redshift dispersion expected is so large that the redshifts remain statistically compatible with conventional ideas about the epoch of earliest galaxy formation. For smaller values of the time dilation signal, e.g. 1.75 or 1.35, even the mean values of the redshifts are in the conventional range. More generally, we have discussed the effects of density evolution and luminosity function on the redshift ranges inferred from an observed flux distribution with various amounts of intrinsic pulse energy stretching for arbitrary values of an observed cosmological time dilation, which may be useful in the interpretation of future experiments.

*Acknowledgments:* We are grateful to M.J. Rees and E. Fenimore for stimulating discussions, I. Horváth and V. Karas and E. Fenimore for useful interactions, and J. Horack for detailed comments on the manuscript. This research has been supported by NASA NAG5-2857. P.M. is grateful for the hospitality of the Institute for Theoretical Physics, University of California, and support through NSF Grant PHY94-07194.

## 7. Appendix: Details of equations (18-19)

One has
$$\binom{n}{k} = \binom{n}{n-k} = \frac{n!}{k!(n-k)!} = \frac{n(n-1)...(n-k+1)}{k!}, \quad n,k = 0,1,2,..., \quad k \leq n.$$

First, we prove the following relation (here $n > 0$):
$$\sum_{k=0}^{n} (-1)^k \binom{n}{k} = 0. \tag{A.1}$$

Proof: If $n = 2x - 1$, where $x$ is natural, one has $(-1)^k = -(-1)^{n-k}$ for any $k \leq n$. Then
$$\binom{n}{k}; \quad \binom{n}{n-k}$$
occur with opposite signs, and they zero give in (A.1). We have $x$ such pairs in the sum of (A.1); hence, here (A.1) holds. If $n = 2x$, where $x$ is natural, then one may write
$$\sum_{k=0}^{2x} (-1)^k \binom{2x}{k} = \binom{2x}{0} + \binom{2x}{2} + .... + \binom{2x}{2x} - \binom{2x}{1} - \binom{2x}{3} - ... - \binom{2x}{2x-1}.$$

We have $(x+1)$ positive terms (the first and the last terms equal to 1), and $x$ negative terms. We will use the following relation:
$$\binom{n}{k} = \binom{n-1}{k} + \binom{n-1}{k-1},$$
where $n > 0$ and $k > 0$. Hence
$$\sum_{k=0}^{2x} (-1)^k \binom{2x}{k} = 2 + \sum_{m=1}^{x-1} \left[\binom{2x-1}{2m} + \binom{2x-1}{2m-1}\right] - \sum_{m=1}^{x} \left[\binom{2x-1}{2m-1} + \binom{2x-1}{2m-2}\right].$$

(The two factors of unity were written down separately; note the sum indices.) After some arrangement it follows that
$$\sum_{k=0}^{2x} (-1)^k \binom{2x}{k} = 2 + \sum_{m=1}^{x-1} \binom{2x-1}{2m} - 1 - \sum_{m=1}^{x} \binom{2x-1}{2m-2} = 1 + \sum_{m=1}^{x-1} \binom{2x-1}{2m} - \sum_{r=0}^{x-1} \binom{2x-1}{2r} = 0,$$
which completes the proof of (A.1).

Using (A.1) it follows that (for $n \geq 2$)
$$\sum_{k=1}^{n} (-1)^k k \binom{n}{k} = n \sum_{k=1}^{n} (-1)^k \binom{n-1}{k-1} = n \sum_{m=0}^{n-1} (-1)^{m+1} \binom{n-1}{m} = 0, \tag{A.2}$$
and ($0 \leq s \leq n-2$):
$$\sum_{k=s+1}^{n} (-1)^k k(k-1)...(k-s) \binom{n}{k} = n(n-1)...(n-s) \sum_{k=s+1}^{n} (-1)^k \binom{n-s-1}{k-s-1} =$$

$$n(n-1)...(n-s) \sum_{m=0}^{n-s-1} (-1)^{m+s+1} \binom{n-s-1}{m} = 0. \tag{A.3}$$

We define $8 - 2D = M$; $M = 0, 1, 2, 3, ...$, and we have ($0 \leq k \leq M$):

$$a_k = \frac{3(-1)^k}{k+3} \binom{M}{k},$$

and thus (if $M \geq 1$):

$$\sum_{k=0}^{M} (k+3) a_k = 0.$$

In what follows we will take $M \geq 2$. The cases $M = 0$ and $M = 1$ will be discussed later. One may write

$$\sum_{k=0}^{M} a_k \frac{y^k}{(1+y)^k} = \sum_{k=0}^{M} a_k \left(1 - \frac{1}{1+y}\right)^k =$$

$$\binom{0}{0} a_0$$

$$+ \binom{1}{0} a_1 - \binom{1}{1} a_1 \frac{1}{1+y}$$

$$+ \binom{2}{0} a_2 - \binom{2}{1} a_2 \frac{1}{1+y} + \binom{2}{2} a_2 \frac{1}{(1+y)^2}$$

$$+ ...$$

$$+ \binom{k}{0} a_k - \binom{k}{1} a_k \frac{1}{1+y} + ... + (-1)^k \binom{k}{k} a_k \frac{1}{(1+y)^k}$$

$$+ ...$$

$$+ \binom{M}{0} a_M - \binom{M}{1} a_M \frac{1}{1+y} + ... + (-1)^M \binom{M}{M} a_M \frac{1}{(1+y)^M}.$$

This "Christmas-tree" arrangement allows one to proceed as follows. First, consider the first terms of the horizontal lines. Their sum is

$$S_0 = \sum_{k=0}^{M} \binom{k}{0} a_k = \sum_{k=0}^{M} a_k = 3 \int_0^1 (1-\chi)^M \chi^2 d\chi = 3 \int_0^1 \eta^M (1-\eta)^2 d\eta =$$

$$\frac{3}{M+1} - \frac{6}{M+2} + \frac{3}{M+3} = \frac{6}{(M+1)(M+2)(M+3)} = A_D.$$

Second, consider the second terms of the horizontal lines. Their sum is

$$S_1 = -\sum_{k=1}^{M} \binom{k}{1} a_k = -\sum_{k=0}^{M} k a_k = -\sum_{k=0}^{M} (k+3-3) a_k = 3 A_D.$$

(The lower index may be changed from 1 to zero, because we add only a vanishing term.) Similarly, we obtain for the third terms

$$S_2 = \sum_{k=2}^{M} \binom{k}{2} a_k = \sum_{k=2}^{M} \frac{k(k-1)}{2!} a_k = \sum_{k=0}^{M} \frac{k(k+3-4)}{2} a_k = \sum_{k=0}^{M} (-2) k a_k = 6 A_D.$$

Consider now the s-th terms ($s \geq 2$). Their sum is:

$$S_{s-1} = (-1)^{s-1} \sum_{k=s-1}^{M} \binom{k}{s-1} a_k = (-1)^{s-1} \sum_{k=0}^{M} \frac{k(k-1)...(k+3-s)(k+2-s)}{(s-1)!} a_k =$$

$$(-1)^{s-1} \sum_{k=0}^{M} \frac{k(k-1)...(k+3-s)(k+3-s-1)}{(s-1)!} a_k =$$

$$\frac{(-1)(-s-1)}{s-1} \sum_{k=0}^{M} (-1)^{s-2} \frac{k(k-1)...(k+3-s)}{(s-2)!} a_k = \frac{s+1}{s-1} S_{s-2}.$$

¿From this the following recurrent relation follows:

$$S_m = \frac{m+2}{m} S_{m-1}; \quad m = 1, 2, ..., (M-1); \quad S_0 = A_D.$$

This may be changed into the direct relation as follows:

$$S_m = \frac{m+2}{m} S_{m-1}, \quad S_{m-1} = \frac{m+1}{m-1} S_{m-2}, \quad ... \quad S_2 = \frac{4}{2} S_1, S_1 = \frac{3}{1} S_0 = 3 A_D.$$

Hence

$$S_m = \frac{(m+2)(m+1)}{2} A_D. \tag{A.4}$$

This relation holds for any $0 \leq m \leq M$.

This "Christmas-tree" proof was done for $M \geq 2$. If $M = 0$, $A_D = 1$ and $S_0 = 1$. If $M = 1$, it is easy to show that $A_D = 1/4$, $S_m = A_D(m+1)(m+2)/2$, $m = 0, 1$. Hence (A.4) holds also for $M = 0$ and $M = 1$.

Using (A.4) one may write

$$\int \frac{t^4 dt}{(1+t^4)^3} \sum_{k=0}^{M} a_k \frac{t^{4k}}{(1+t^4)^k} = A_D \int \frac{t^4 dt}{(1+t^4)^3} \sum_{k=0}^{M} \frac{(k+1)(k+2)}{2(1+t^4)^k} = A_D \psi_M(t). \tag{A.5}$$

To do the concrete integration one may proceed as follows ($M = 0, 1, 2, ...$). We write:

$$\psi_0(t) = \int \frac{t^4 dt}{(1+t^4)^3}.$$

$$\psi_1(t) = \frac{t^5}{4(1+t^4)^3} + \frac{33}{12} \psi_0(t).$$

The function $\psi_0(t)$ is a tabulated integral. Consider now $M \geq 2$, and assume that we know $\psi_{M-1}(t)$ and $\psi_{M-2}(t)$. Then one may write:

$$\psi_M(t) - \psi_{M-1}(t) = \frac{(M+1)(M+2)}{2} \int \frac{t^4 dt}{(1+t^4)^{M+3}}$$

$$= \frac{(M+1)t^5}{8(1+t^4)^{M+2}} + \frac{4M+3}{4M}\Big(\psi_{M-1}(t) - \psi_{M-2}(t)\Big).$$

Table 1

| $D$ | $\mathcal{L}_o/(10^{57}h^{-2})$ | $\overline{(1+z_b)}$ | $\Delta(1+z_b)$ | $\overline{(1+z_{d'})}$ | $\Delta(1+z_{d'})$ | $r_{d'b}$ | $\Delta r_{d'b}$ |
|---|---|---|---|---|---|---|---|
| 1.0 | 0.2 | 1.15 | 0.036 | 1.59 | 0.09 | 1.39 | 0.09 |
| 1.5 | 0.3 | 1.18 | 0.046 | 1.74 | 0.11 | 1.47 | 0.11 |
| 2.0 | 0.45 | 1.23 | 0.057 | 1.94 | 0.14 | 1.58 | 0.14 |
| 2.5 | 0.6 | 1.26 | 0.066 | 2.11 | 0.17 | 1.67 | 0.16 |
| 3.0 | 0.8 | 1.31 | 0.078 | 2.32 | 0.21 | 1.78 | 0.19 |
| 3.5 | 1.1 | 1.36 | 0.093 | 2.61 | 0.26 | 1.91 | 0.23 |
| 4.0 | 2.0 | 1.50 | 0.132 | 3.36 | 0.40 | 2.23 | 0.33 |
| 4.5 | 7.0 | 2.02 | 0.287 | 6.55 | 1.04 | 3.24 | 0.69 |
| 5.0 | 15.0 | 2.63 | 0.474 | 10.90 | 1.96 | 4.14 | 1.05 |

*Table 1 Caption: Standard Candle results based on equations (38, 39) of the mean $(1+z)$ values and their ratio and dispersions for different $D$, where the different $\mathcal{L}_o$ (in units photons/($cm^2 s$)) used for different $D$ are taken from the fits of HMM95.*

Table 2

| $D$ | $\mathcal{L}_o/(10^{57}h^{-2})$ | $\overline{(1+z_b)}$ | $\Delta(1+z_b)$ | $\overline{(1+z_{d'})}$ | $\Delta(1+z_{d'})$ | $r_{d'b}$ | $\Delta r_{d'b}$ |
|---|---|---|---|---|---|---|---|
| 4.5 | 5.0 | 1.84 | 0.23 | 5.38 | 0.80 | 2.92 | 0.57 |
| 4.5 | 6.0 | 1.93 | 0.26 | 5.98 | 0.92 | 3.09 | 0.63 |
| 4.5 | 7.0 | 2.02 | 0.29 | 6.55 | 1.04 | 3.24 | 0.69 |
| 4.5 | 8.0 | 2.10 | 0.31 | 7.11 | 1.16 | 3.38 | 0.75 |
| 4.5 | 9.0 | 2.18 | 0.34 | 7.65 | 1.28 | 3.51 | 0.80 |

*Table 2 Caption: Standard Candle results showing the effects on the mean redshifts of allowing the value of $\mathcal{L}_o$ (ph cm$^{-2}$ s$^{-1}$) to vary within $\lesssim 1s\sigma$ of its best fit value for fixed $D$.*

Table 3

| $D$ | $\mathcal{L}_m/(10^{57}h^{-2})$ | $\overline{(1+z_b)}$ | $\Delta(1+z_b)$ | $\overline{(1+z_{d'})}$ | $\Delta(1+z_{d'})$ | $r_{d'b}$ | $\Delta r_{d'b}$ |
|---|---|---|---|---|---|---|---|
| 2.00 | 0.05 | 1.33 | 0.22 | 1.82 | 0.63 | 1.37 | 0.52 |
| 2.50 | 0.08 | 1.42 | 0.28 | 2.07 | 0.86 | 1.46 | 0.67 |
| 3.00 | 0.10 | 1.49 | 0.33 | 2.33 | 1.12 | 1.57 | 0.83 |
| 3.50 | 0.15 | 1.62 | 0.42 | 2.83 | 1.63 | 1.75 | 1.11 |
| 4.00 | 0.30 | 1.92 | 0.66 | 4.06 | 2.99 | 2.12 | 1.72 |
| 4.50 | 0.50 | 2.32 | 0.98 | 6.29 | 5.34 | 2.71 | 2.57 |
| 5.00 | 0.80 | 2.96 | 1.46 | 11.00 | 9.63 | 3.70 | 3.73 |

*Table 3 Caption: Power Law luminosity distribution results showing average redshifts calculated from equations (38, 39), notation as in Table 1. A ratio of $K = \mathcal{L}_M/\mathcal{L}_m = 100$ was taken everywhere, and values of $\mathcal{L}_m$ (ph cm$^{-2}$ s$^{-1}$) for each $D$ are taken from the best fits of HMM95.*

Table 4

| $D$ | $\mathcal{L}_m/(10^{57}h^{-2})$ | $\overline{(1+z_b)}$ | $\Delta(1+z_b)$ | $\overline{(1+z_{d'})}$ | $\Delta(1+z_{d'})$ | $r_{d'b}$ | $\Delta r_{d'b}$ |
|---|---|---|---|---|---|---|---|
| 4.50 | 0.20 | 1.84 | 0.56 | 4.22 | 2.82 | 2.29 | 1.69 |
| 4.50 | 0.40 | 2.18 | 0.85 | 5.68 | 4.56 | 2.60 | 2.32 |
| 4.50 | 0.60 | 2.44 | 1.09 | 6.86 | 6.08 | 2.81 | 2.79 |
| 4.50 | 0.80 | 2.66 | 1.31 | 7.91 | 7.48 | 2.97 | 3.17 |
| 4.50 | 1.00 | 2.86 | 1.51 | 8.87 | 8.81 | 3.11 | 3.49 |

*Table 4 Caption: Mean redshifts for the power law luminosity function; $D$ and $K$ are fixed; $K = \mathcal{L}_M/\mathcal{L}_m = 100$ and $\mathcal{L}_m$ (ph cm$^{-2}$ s$^{-1}$) is varied.*

Table 5

| $D$ | $\mathcal{L}_m/(10^{57}h^{-2})$ | $K$ | $\overline{(1+z_b)}$ | $\Delta(1+z_b)$ | $\overline{(1+z_{d'})}$ | $\Delta(1+z_{d'})$ | $r_{d'b}$ | $\Delta r_{d'b}$ |
|---|---|---|---|---|---|---|---|---|
| 4.5 | 0.5 | 4 | 1.37 | 0.12 | 2.61 | 0.47 | 1.90 | 0.38 |
| 4.5 | 0.5 | 64 | 2.10 | 0.74 | 5.51 | 3.95 | 2.63 | 2.10 |
| 4.5 | 0.5 | 324 | 3.12 | 2.04 | 8.88 | 11.4 | 2.85 | 4.10 |
| 4.5 | 0.5 | 1024 | 4.29 | 4.14 | 12.2 | 23.0 | 2.84 | 6.03 |
| 4.5 | 0.5 | 2500 | 5.51 | 7.11 | 15.3 | 39.1 | 2.77 | 7.94 |

*Table 5 Caption: Mean redshift in the case of a power law distribution. $\mathcal{L}_m$ (ph cm$^{-2}$ s$^{-1}$) and $D$ are fixed, and $K = \mathcal{L}_M/\mathcal{L}_m$ varies.*

Table 6

| $\Omega$ | 1.0 | 0.8 | 0.6 | 0.4 | 0.2 |
|---|---|---|---|---|---|
| $z_b$ | 0.6 | 0.7 | 0.8 | 1.1 | 2.0 |
| $z_{d'}$ | 5.2 | 5.5 | 6.0 | 7.0 | 10.6 |

*Table 6 Caption: Values of the mean resdhifts derived for various values of $\Omega$.*

**Figure Captions**

*Figure 1.* Standard candle (SC) theoretical integral peak flux distributions for values $D = 2, 3, 4, 5, 6, 7$ in increasing order upwards. Two sets of curves are shown, for two values of the luminosity $\mathcal{L}_o$. The bottom set has a luminosity $\mathcal{L}_o = 4.5 \times 10^{56} h^{-2}$ s$^{-1}$, the upper one has $\mathcal{L}_o = 3 \times 10^{57} h^{-2}$ s$^{-1}$. The lower set has been artifically moved down by one unit in $\log N$ so as not to overlap. For comparison, the 2B data set is also shown twice, again the lower curve down by one unit, to illustrate qualitatively that different choices of $\mathcal{L}_o$ provide approximate fis to the same data set corresponding to different evolution indices $D$. The higher luminosities fit with higher $D$. In this case the lower curves fit approximately to $D = 2$, the upper ones with $D = 4$ (and similar fits can be found for $D = 3$. More accurate $\chi^2$ fits are discussed in HMM95, see also Table 1.

*Figure 2.* Power law luminosity function (PL) theoretical integral peak flux distribution for $D = 2, 3, 4, 5, 6, 7$ increasing upward, for $\beta = 15/8$. Two sets of curves are shown, both for $K = \mathcal{L}_M/\mathcal{L}_m = 10^2$. The lower set (which has been offset downwards by one unit) is for $(\mathcal{L}_M, \mathcal{L}_m) = (5 \times 10^{57} h^{-2}$ s$^{-1}, 5 \times 10^{55} h^{-2}$ s$^{-1})$ and suggests a fit for $D = 2$, while the upper set has $(\mathcal{L}_M, \mathcal{L}_m) = (2.5 \times 10^{58} h^{-2}$ s$^{-1}, 2.5 \times 10^{56} h^{-2}$ s$^{-1})$ and suggests a fit to $D = 4$, when compared to the 2B data set. Again, higher luminosities provide qualitative fits to higher evolution indices $D$. For more detailed fits, see HMM95 and Table 3).

*Figure 3.* A comparison of theoretical integral peak flux distributions for standard candle (SC, upper curves) and power law (PL, lower curves) luminosity functions. The curves in each set are, in upward order, for density evolution indices $D = 2, 3, 4, 5, 6, 7$. Arbitrary values of $\mathcal{L}_o = 10^{59} h^{-2}$ s$^{-1}$ for SC and $(\mathcal{L}_M, \mathcal{L}_m, K) = (10^{59} h^{-2}$ s$^{-1}, 10^{54} h^{-2}$ s$^{-1}, 10^5)$ for PL cases were chosen. Both SC and PL show (for $D \leq 4$) two regimes, $N(> F) \propto F^{-3/2}$ at high $F$ and $F \to$ constant at low $F$. The large ratio $K = \mathcal{L}_M/\mathcal{L}_m$ in the PL case was chosen to illustrate the third asymptotic regime in this case at intermediate $F$, namely $N(> F) \propto F^{1-\beta}$, where $\beta$ is the index of the power law luminosity function.

*Figure 4.* Standard candle luminosity function: Mean $(1 + z)$ for dim+dimmest (bottom curve) and bright (upper curve) bursts, and their ratio $r_{d'b}$ (middle curve), as a function of the density evolution parameter $D$. Values are computed for integer and half-integer values of $D$, but are drawn slightly offset in the figure so the error bars can be distinguished. For a time dilation ratio $\tau_{d'b}$ one has a redshift factor ratio $r_{d'b} = \tau_{d'b}$ ($r_{d'b} = \tau_{d'b}^{5/3}$) if one neglects (includes) energy stretching of time profiles. Results based on equations (38, 39), as in Table 1, where the different $\mathcal{L}_o$ used for different $D$ are taken from the SC fits of HMM95.

*Figure 5.* Power Law luminosity distribution: Mean $(1 + z)$ for dim+dimmest (bottom curve) and bright (upper curve) bursts, and their ratio $r_{d'b}$ (middle curve), as a function of the density evolution parameter $D$. Values are computed for integer and half-integer values of $D$, but are drawn slightly offset in the figure so the error bars can be distinguished. For a time dilation ratio $\tau_{d'b}$ one has a redshift factor ratio $r_{d'b} = \tau_{d'b}$ ($r_{d'b} = \tau_{d'b}^{5/3}$) if one neglects (includes) energy stretching of time profiles. The results are calculated from equations (38, 39), as in Table 3, using

a ratio of maximum/minimum luminosities $K = 100$ and values of $\mathcal{L}_m$ for each $D$ taken from the PL fits of HMM95.

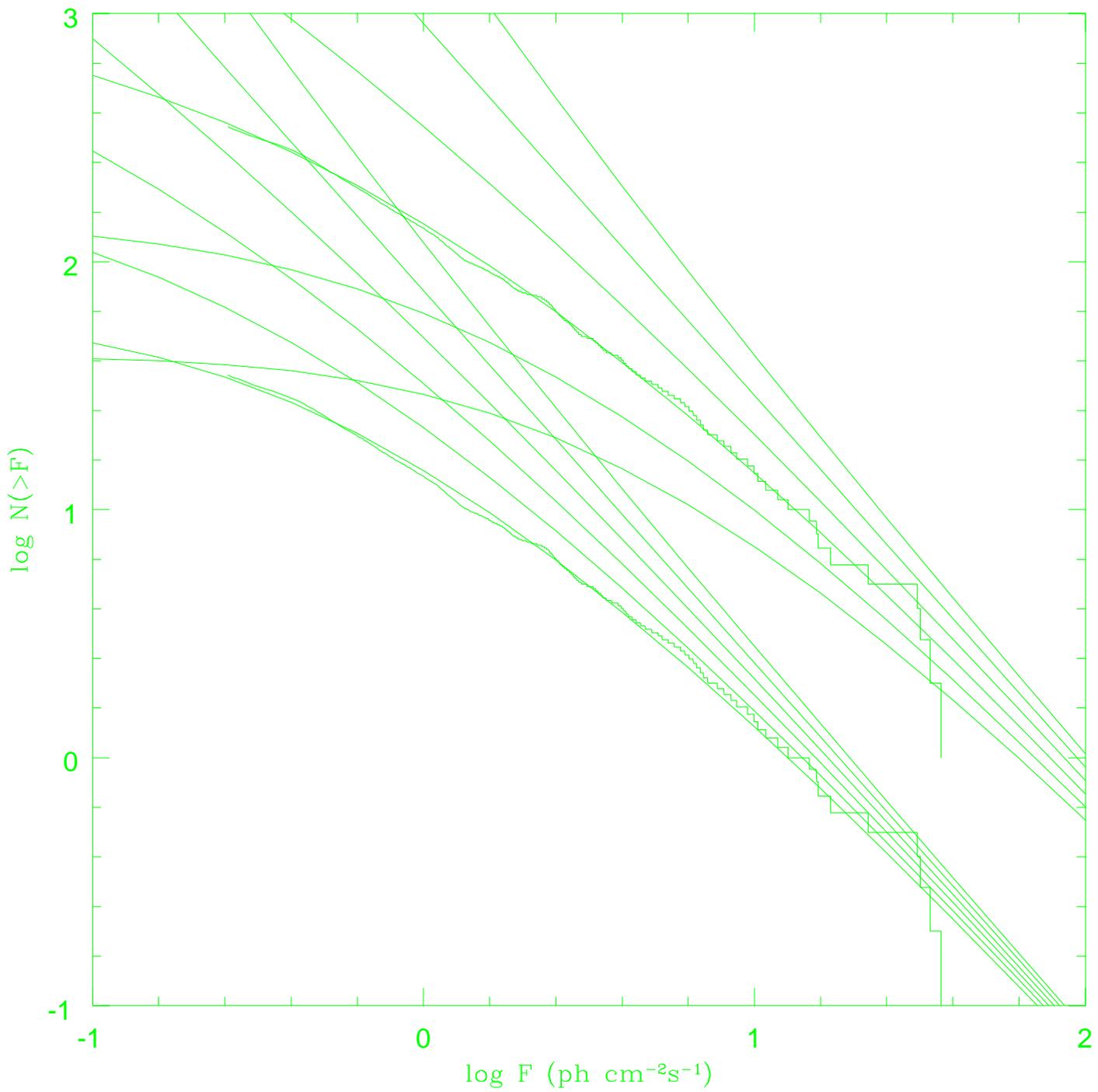

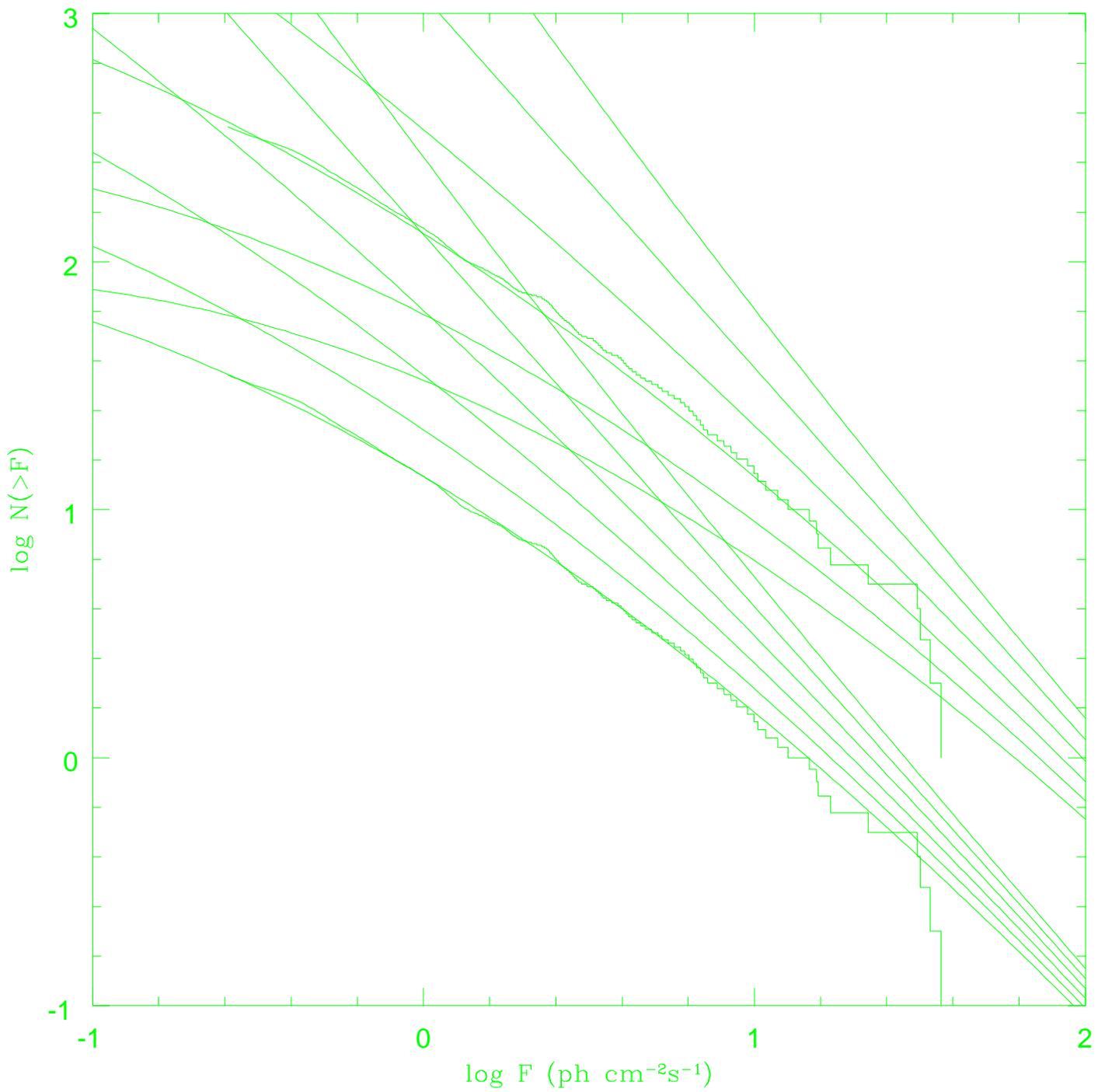

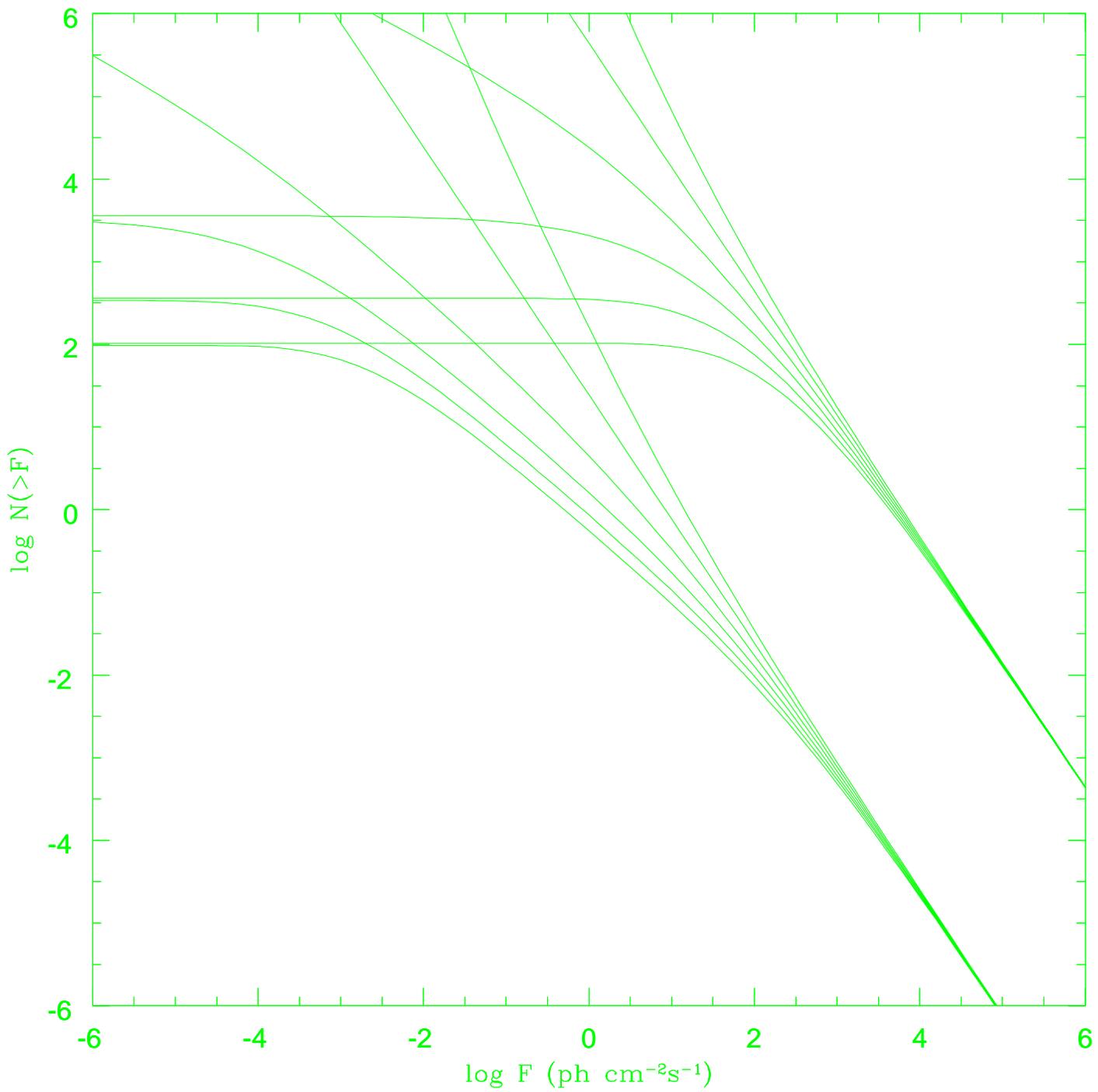

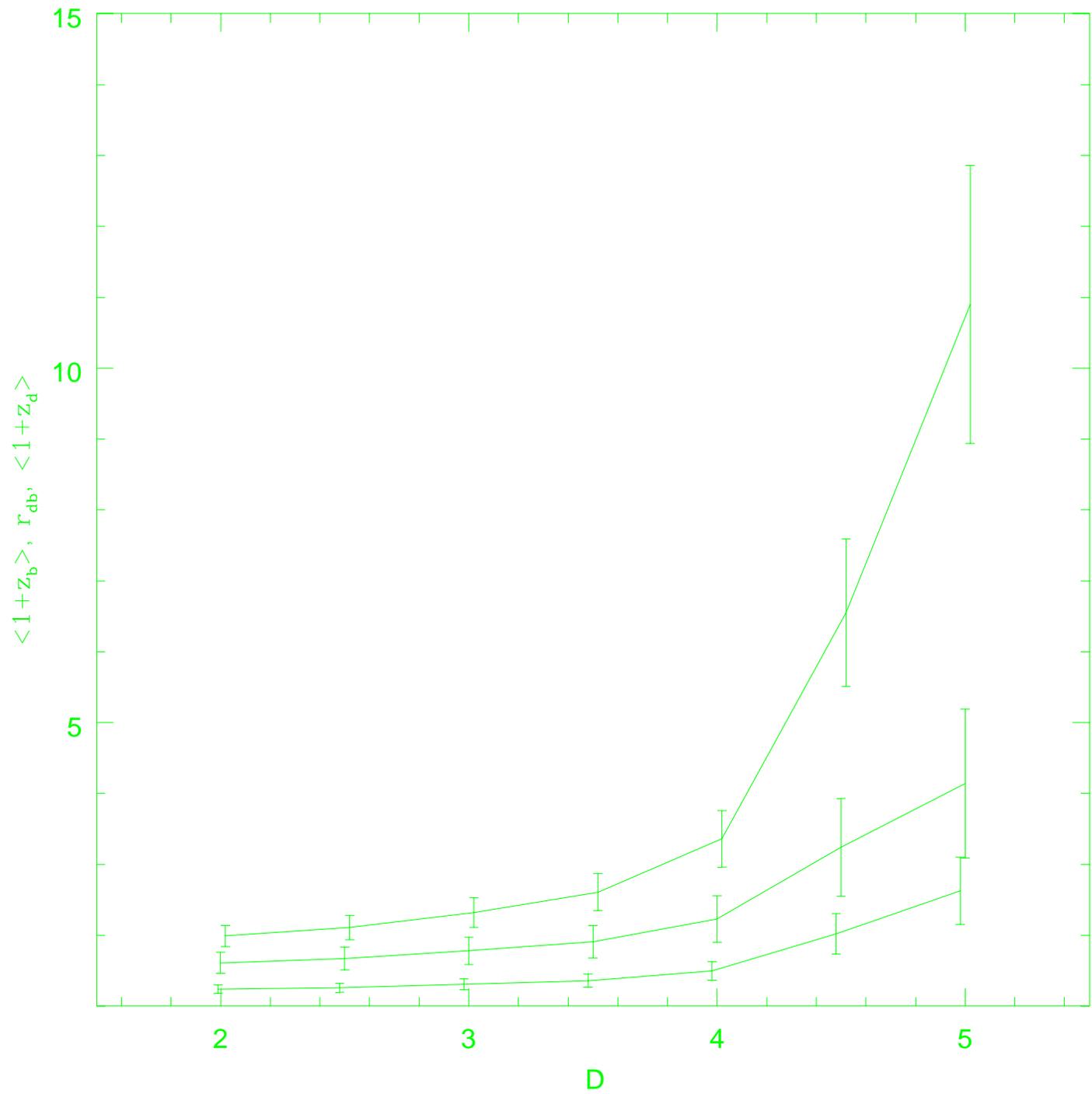

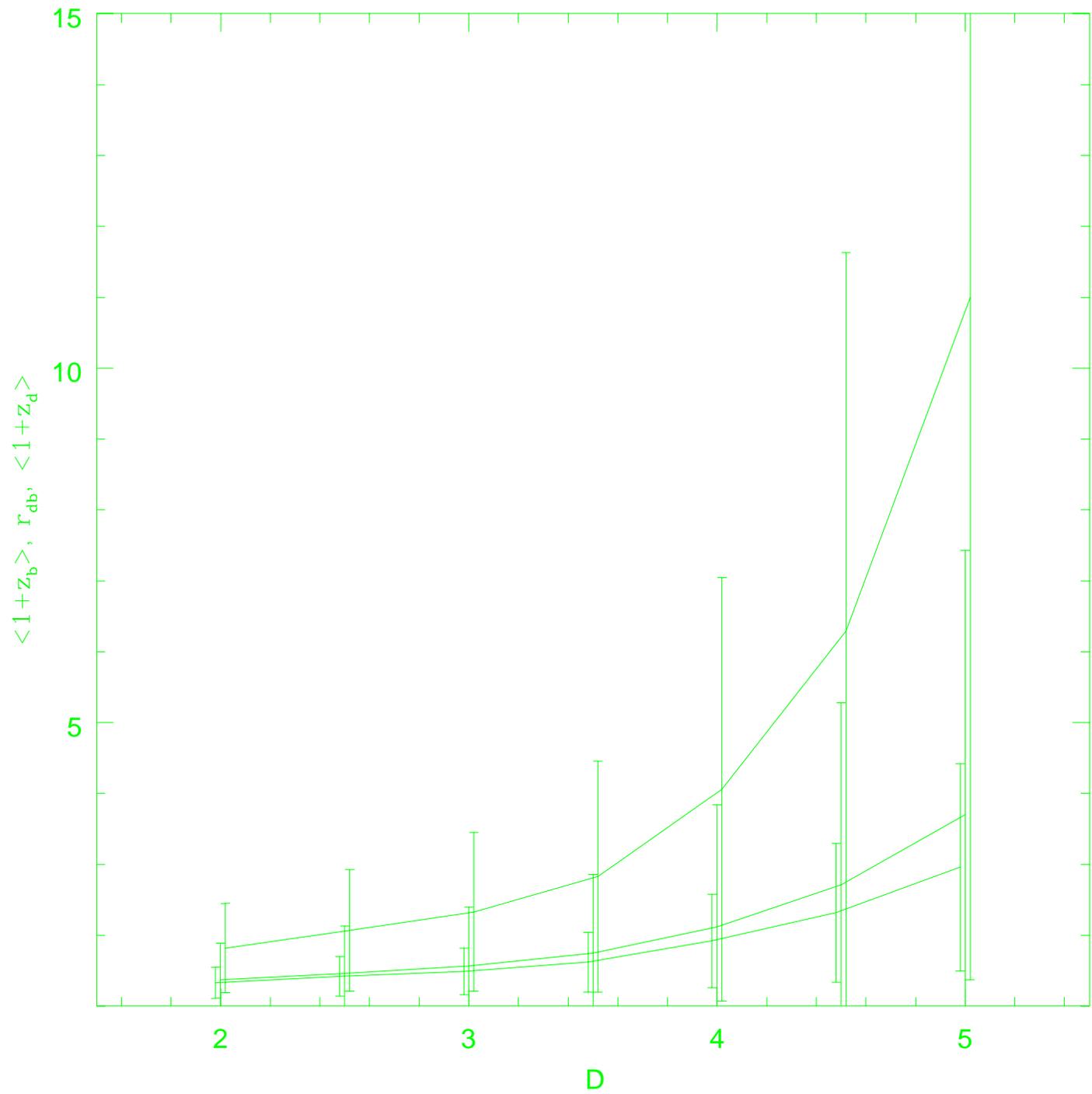